\documentclass[a4paper,11pt]{article}
\usepackage{jcappub} 
\usepackage{lineno}

\usepackage{dcolumn}
\usepackage{verbatim}

\usepackage{breakurl}

\usepackage{lmodern}

\pdfoutput=1
\usepackage{graphicx}
\usepackage{bm}
\usepackage{amsmath}
\usepackage{amsfonts}
\usepackage{amssymb}
\usepackage{multirow}
\usepackage[capitalise]{cleveref}
\usepackage{acro}
\usepackage[utf8]{inputenc}

\usepackage{tikz}
\usepackage{pgfplots}

\crefname{figure}{Fig.}{Figs.}

\newcommand{\be}{{\begin{eqnarray}}}
\newcommand{\ee}{{\end{eqnarray}}}

\newcommand{\overbar}[1]{\mkern 1.5mu\overline{\mkern-1.5mu#1\mkern-1.5mu}\mkern 1.5mu}

\newcommand{\cA}{\mathcal{A}}
\newcommand{\cI}{\mathcal{I}}
\newcommand{\cT}{\mathcal{T}}
\newcommand{\cP}{\mathcal{P}}

\newcommand{\cH}{\mathcal{H}}
\newcommand{\cS}{\mathcal{S}}
\newcommand{\cO}{\mathcal{O}}

\newcommand{\bk}{\mathbf{k}}

\newcommand{\bq}{\mathbf{q}}

\newcommand{\bx}{\mathbf{x}}

\newcommand{\ud}{\mathrm{d}}

\newcommand{\Beq}{\begin{align}}
\newcommand{\Eeq}{\end{align}}

\DeclareAcronym{GW}{
  short = GW,
  long = gravitational wave ,
  short-plural = s ,
}
\DeclareAcronym{LIGO}{
  short =LIGO ,
  long = Laser Interferometer Gravitational-Wave Observatory ,
  short-plural = ,
}
\DeclareAcronym{LVK}{
  short = LVK ,
  long = {LIGO, Virgo, and KAGRA},
  short-plural = ,
}

\DeclareAcronym{SGWB}{
  short = SGWB ,
  long = stochastic gravitational-wave background ,
  short-plural = s ,
}
\DeclareAcronym{GWB}{
  short = GWB ,
  long = gravitational-wave background ,
  short-plural = s ,
}
\DeclareAcronym{CBC}{
  short = CBC ,
  long = compact binary coalescence ,
  short-plural = s ,
}
\DeclareAcronym{BH}{
  short = BH ,
  long = black hole ,
  short-plural = s ,
}
\DeclareAcronym{BBH}{
  short = BBH ,
  long = binary black hole ,
  short-plural = s ,
}
\DeclareAcronym{PBH}{
  short = PBH ,
  long = primordial black hole ,
  short-plural = s ,
}
\DeclareAcronym{SNR}{
  short = SNR ,
  long = signal-to-noise ratio ,
  short-plural = s ,
}
\DeclareAcronym{IMRPPv2}{
  short = ,
  long = {\normalsize IMRP}{\footnotesize HENOM}{\normalsize P}v2 ,
  short-plural = ,
}
\DeclareAcronym{PTA}{
  short = PTA ,
  long = pulsar timing array ,
  short-plural = s ,
}
\DeclareAcronym{SFR}{
  short = SFR ,
  long = star formation rate ,
  short-plural =  ,
}
\DeclareAcronym{FRW}{
  short = FRW ,
  long = Friedmann-Robertson-Walker ,
  short-plural =  ,
}
\DeclareAcronym{IMR}{
  short = IMR ,
  long = inspiral-merger-ringdown ,
  short-plural =  ,
}
\DeclareAcronym{LISA}{
	short = LISA ,
	long  = Laser Interferometer Space Antenna,
  short-plural =  ,
}
\DeclareAcronym{ET}{
	short = ET ,
	long  = Einstein Telescope,
  short-plural =  ,
}
\DeclareAcronym{CE}{
	short = CE ,
	long  = Cosmic Explorer,
  short-plural =  ,
}
\DeclareAcronym{BBO}{
	short = BBO ,
	long  = Big Bang Observer,
  short-plural =  ,
}
\DeclareAcronym{DECIGO}{
	short = DECIGO ,
	long  = Deci-hertz Interferometer Gravitational Wave Observatory,
  short-plural =  ,
}
\DeclareAcronym{ABH}{
	short = ABH ,
	long  = astrophysical black hole,
  short-plural = s ,
}

\DeclareAcronym{PNG}{
	short = PNG ,
	long  = primordial non-Gaussianity ,
  short-plural =  ,
}

\DeclareAcronym{CMB}{
	short = CMB ,
	long  = cosmic microwave background ,
  short-plural =  ,
}
\DeclareAcronym{LSS}{
	short = LSS ,
	long  = large-scale structure ,
  short-plural =  ,
}

\DeclareAcronym{PGW}{
	short = PGW ,
	long  = primordial gravitational wave ,
  short-plural = s ,
}
\DeclareAcronym{SIGW}{
	short = SIGW ,
	long  = scalar-induced gravitational wave ,
  short-plural = s ,
}

\DeclareAcronym{IGW}{
	short = IGW ,
	long  = induced gravitational wave ,
  short-plural = s ,
}

\DeclareAcronym{RD}{
	short = RD,
	long  = radiation-dominated ,
  short-plural =  ,
}
\DeclareAcronym{eRD}{
	short = eRD,
	long  = early radiation-dominated ,
  short-plural =  ,
}
\DeclareAcronym{MD}{
	short = MD,
	long  = matter-dominated ,
  short-plural =  ,
}
\DeclareAcronym{eMD}{
	short = eMD,
	long  = early matter-dominated ,
  short-plural =  ,
}
\DeclareAcronym{SW}{
	short = SW,
	long  = Sachs-Wolfe ,
  short-plural =  ,
}
\DeclareAcronym{ISW}{
	short = ISW,
	long  = integrated Sachs-Wolfe ,
  short-plural =  ,
}

\DeclareAcronym{DM}{
	short = DM,
	long  = dark matter ,
  short-plural =  ,
}

\DeclareAcronym{NANOGrav}{
	short = NANOGrav ,
	long  = North American Nanohertz Observatory for Gravitational Waves ,
  short-plural =  ,
}

\DeclareAcronym{EPTA}{
	short = EPTA ,
	long  = European Pulsar Timing Array ,
  short-plural =  ,
}

\DeclareAcronym{PPTA}{
	short = PPTA ,
	long  = Parkes Pulsar Timing Array ,
  short-plural =  ,
}

\DeclareAcronym{CPTA}{
	short = CPTA ,
	long  = Chinese Pulsar Timing Array ,
  short-plural =  ,
}

\DeclareAcronym{PDF}{
	short = PDF ,
	long  = probability distribution function ,
  short-plural = s ,
}

\DeclareAcronym{SMBH}{
  short = SMBH ,
  long  = supper-massive black hole ,
  short-plural = s ,
}
\DeclareAcronym{SKA}{
	short = SKA ,
	long  = Square Kilometre Array,
  short-plural =  ,
}
\DeclareAcronym{NG15}{
  short = NG15 ,    
  long  = NANOGrav 15-year ,
  short-plural =  ,
}
\DeclareAcronym{SM}{
  short = SM ,    
  long  = Standard Model ,
  short-plural =  ,
}
\DeclareAcronym{BSM}{
  short = BSM ,    
  long  = beyond the Standard Model ,
  short-plural =  ,
}
\DeclareAcronym{GUT}{
  short = GUT ,
  long =  grand unified theories
 ,
  short-plural = s ,
}

\title{\boldmath Comprehensive analysis of dissipative effects in the induced gravitational waves}

\author[a,b]{Yan-Heng Yu } 
\emailAdd{yhyu@ihep.ac.cn}
\author[a,b]{, Zhe Chang }
\emailAdd{changz@ihep.ac.cn}
\author[c,\ast]{, Sai Wang \note[$\ast$]{Corresponding author.}}  
\emailAdd{wangsai@hznu.edu.cn}

\affiliation[a]{Theoretical Physics Division, Institute of High Energy Physics, Chinese Academy of Sciences, 19B Yuquan Road, Shijingshan District, Beijing 100049, China}
\affiliation[b]{School of Physical Sciences, University of Chinese Academy of Sciences, 19A Yuquan Road, Shijingshan District, Beijing 100049, China}
\affiliation[c]{School of Physics, Hangzhou Normal University, No.2318 Yuhangtang Road, Yuhang District, Hangzhou 311121, China}

\abstract{Dissipation is an intrinsic property of the cosmic fluid, leading to the damping of curvature perturbations at small scales. In this paper, we comprehensively study dissipative effects in gravitational waves induced by curvature perturbations, known as induced gravitational waves (IGWs). We find dissipative effects become especially significant at wavenumber $k \sim k_{\mathcal{H},\mathrm{dec}}$, where $k_{\mathcal{H},\mathrm{dec}}$ corresponds to the horizon scale at the decoupling of weakly-interacting particles. They can leave characteristic features on the IGW spectrum, including a notable suppression with a ``double-valley'' structure at $k \sim k_{\mathcal{H},\mathrm{dec}}$ and a modified infrared behavior without logarithmic running at $k \lesssim k_{\mathcal{H},\mathrm{dec}}$. Within the Standard Model of particle physics, dissipative effects caused by neutrinos at the nanohertz frequencies can be important in the analysis of pulsar timing array data. Furthermore, dissipation-induced features associated with possible new weakly-interacting particles can be detectable by a wide range of gravitational-wave experiments, serving as a promising probe of new physics at extremely high energy scales. As an extension, we also discuss dissipative effects in the presence of primordial non-Gaussianity and their impacts on the anisotropies of IGWs and the poltergeist mechanism. These dissipative effects not only provide a more realistic description of IGWs but also exhibit rich phenomenology and profound physical implications, opening a new window into understanding the early Universe and fundamental physics. }

\begin{document}

\maketitle
\flushbottom
\allowdisplaybreaks

\section{Introduction}\label{sec:1}

Understanding the early Universe is one of the central objectives for cosmology and fundamental physics. 
\Acp{GW}, due to their nature of hardly interacting with other matters, provide a unique window to detect the Universe's earliest epochs, which are otherwise inaccessible to electromagnetic probes. 
Cosmological \acp{SGWB}, including those form inflationary quantum fluctuations, preheating, phase transitions, topological defects, and cosmological scalar perturbations, are searched across a wide range of frequency bands.
To be specific, ground-based \ac{GW} interferometers like \ac{LVK} \cite{Harry_2010,VIRGO:2014yos,Somiya:2011np} have hunted for the hundred-hertz \ac{SGWB} and established upper limits \cite{LIGOScientific:2025bgj}, space-based \ac{GW} interferometers like the upcoming \ac{LISA} \cite{LISA:2024hlh} are anticipated to seek the milli-hertz \ac{SGWB} in the next decade, and, most excitingly, \ac{PTA} experiments (i.e., \ac{NANOGrav} \cite{Jenet:2009hk}, \ac{EPTA} \cite{Kramer:2013kea}, \ac{PPTA} \cite{Manchester:2012za}, and Chinese Pulsar Timing Array (CPTA) \cite{2016ASPC..502...19L}) have already found strong evidences of the nano-hertz \ac{SGWB} \cite{NANOGrav:2023gor,EPTA:2023fyk,Reardon:2023gzh,Xu:2023wog}.
These cosmological \acp{SGWB}, once detected, will bring us a wealth of physical information about the early Universe, marking a milestone to explore cosmology and fundamental physics.

\Acp{IGW} \cite{Mollerach:2003nq,Ananda:2006af,Baumann:2007zm,Assadullahi:2009jc,Espinosa:2018eve,Kohri:2018awv,Domenech:2021ztg}, as an important constituent of the cosmological \acp{SGWB}, have garnered substantial attention recently.
They exhibit close connections to many crucial problems in the early Universe, e.g., inflation, cosmic evolution, \acp{PBH}, etc., and can also well explain the \ac{SGWB} reported by \acp{PTA} \cite{NANOGrav:2023hvm}.
These \acp{IGW} are naturally predicted by the cosmological perturbation theory, referring to the secondary \acp{GW} induced by the nonlinear couplings of linear cosmological scalar perturbations $\phi$.
The production of \acp{IGW} is determined by both the primordial curvature power spectrum $\cP_\zeta$ and the evolution of $\phi$ after inflation. 
While \ac{CMB} observations set $\cP_\zeta\sim 10^{-9}$ on large scales \cite{Planck:2018vyg}, $\cP_\zeta$ is allowed to be large enough to produce observable \acp{IGW} on small scales.
Particularly, an enhanced small-scale $\cP_\zeta$ is also well-motivated by the formation of \acp{PBH}, which are one of the leading dark matter candidates.
For a given $\cP_\zeta$, the \acp{IGW} are then determined by the classical evolution of $\zeta$ after horizon reentry, which highly depends on the properties of cosmic fluid. 
Therefore, the \acp{IGW} also serve as a sensitive tracer of these properties of cosmic fluid in the early Universe.

The focus of this paper is to discuss the relationship between the \acp{IGW} and an inherent nature of cosmic fluid, namely, the dissipation in it. 
In most existing studies, the cosmic fluid is simply assumed as a perfect fluid without dissipation. 
This can be a good approximation if the mean free path of particles in cosmic fluid is negligible, leading to a simplified analysis of the evolution of $\zeta$.
Based on this assumption, we can establish direct links between the \acp{IGW} and the early Universe.
The energy-density spectrum of \acp{IGW}, $\Omega_\mathrm{gw}$, in the standard \ac{RD} Universe can be calculated through a semi-analytical method \cite{Espinosa:2018eve,Kohri:2018awv}. 
Besides, other crucial issues, such as non-standard cosmic expansion history, primordial non-Gaussianity, and \ac{GW} anisotropies, have also been widely investigated in the study of \acp{IGW}, providing deeper insights into the early Universe.
So far so good.

However, since the mean free path of particles is non-vanishing in reality, the cosmic fluid essentially has viscosity and thermal conduction, with inevitable dissipative effects in it. 
In fact, the diffusion of particles from high-density to low-density regions can erase the sound waves in cosmic fluid on scales smaller than a typical diffusion scale $k_D^{-1}$.
As a result, these sound waves for the modes $k\gg k_D$ dissipate their energy with an exponential damping \cite{Weinberg:1971mx}.
These dissipative effects, as a universal property of fluid, have important applications in cosmology.
The damping in the photon-baryon plasma caused by the photon diffusion is well-known as “Silk damping” \cite{Silk:1967kq}, which results in the suppression of the \ac{CMB} angular power spectrum at the multipoles $\ell>10^3$.
Similar dissipative effects due to the particle diffusion generally exist in the earlier Universe, offering valuable insights into small-scale $\zeta$ \cite{Jeong:2014gna}.
As for \acp{IGW}, since dissipative effects can significantly affect the evolution of $\zeta$, they should be a basic factor to consider in the study of \acp{IGW}.

Though of great importance, dissipative effects in \acp{IGW} have long been ignored until recent works.
Ref.~\cite{Yu:2024xmz} firstly investigated dissipative effects in \acp{IGW} and also named them as ``Silk damping''.
The main effect of dissipation (for a broad $\cP_\zeta$) is suppressing $\Omega_\mathrm{gw}$ at the frequencies related to the decoupling of weakly-interacting particles.
By revealing the relation between these frequencies and particle interactions, Ref.~\cite{Yu:2024xmz} proposed that dissipative effects in \acp{IGW} can be a novel probe to new physics, as an essential complement to particle physics methods.
In Ref.~\cite{Domenech:2025bvr}, the authors concentrated more on the case of a sharply-peaked $\cP_\zeta$.
In this case, the dissipation modifies $\Omega_\mathrm{gw}$ by regularizing the divergent peak at the resonant frequency and removing the typical logarithmic running in the infrared region. 
Additionally, semi-analytical methods for some special conditions are developed to calculate $\Omega_\mathrm{gw}$.

In this paper, we will present a comprehensive and in-depth study of dissipative effects in \acp{IGW}.  
We will start from the particle interactions in the early universe and establish their general relationship with the dissipative effects. 
In more general cases (different types of $\cP_\zeta$ and various dissipation parameters), we will conduct a detailed study on how dissipation influences the \ac{IGW} spectrum.
Our results uncover a rich phenomenology of dissipative effects.
Notably, we newly find that dissipation can produce a characteristic multi-valley structure and induce complex modifications to the infrared behavior.
Furthermore, we will provide an analysis of the physical origins behind these phenomena.  
As an extension, we will also explore the interplay between dissipative effects and primordial non-Gaussianity, \ac{GW} anisotropies, and the poltergeist mechanism.  
The implications of our work include:  
(i) Since $\Omega_{\mathrm{gw}}$ is the most basic observable for \acp{IGW}, its precise calculation is essential for \ac{GW} data analysis. 
Accounting for the universal and significant role of dissipation, our study offers a more realistic description of \acp{IGW}, making it highly important for upcoming \acp{GW} experiments. 
(ii) Identifying the sources of \ac{SGWB} is crucial but challenging. 
The dissipation-induced feature on $\Omega_{\mathrm{gw}}$ offer a powerful way to differentiate \acp{IGW} from other \ac{GW} origins.  
Especially, dissipative effects from neutrinos are expected to provide insightful insights into whether \acp{IGW} can account for the nano-hertz \ac{SGWB} reported by \acp{PTA}.  
(iii) Dissipative effects in \acp{IGW}, especially the newly discovered multi-valley feature, could act as unique signatures of particle decoupling in the early Universe.  
This can opens the door to extracting valuable insights into fundamental physics at extremely high energy scales.

This paper is organized as follows.
In Sec.~\ref{sec:2}, we establish the theoretical foundation of this work by deriving $\Omega_\mathrm{gw}$ in the presence of dissipation.
In Sec.~\ref{sec:3}, we demonstrate the numerical results of $\Omega_\mathrm{gw}$ for different types of $\cP_\zeta$. 
We also summarize the properties of dissipatives effects and analyze their physical origins. 
In Sec.~\ref{sec:4}, we emphasize physical implications of these dissipative effects in \acp{IGW} for \ac{GW} observations and for new physics searches.
In Sec.~\ref{sec:5}, we discuss the dissipative effects in the presence of primordial non-Gaussianity, and how dissiaption affects anisotropies of \acp{IGW} and the poltergeist mechanism.   
Sec.~\ref{sec:6} makes a conclusion of our work.

\section{Theoretical foundations}\label{sec:2}

\subsection{Dissipation in cosmic fluid}\label{sec:2.1}

We start from considering the hydrodynamic properties of cosmic fluid in the \ac{RD} Universe.
Due to the non-vanishing mean free path of particles, the cosmic plasma should be described as an imperfect relativistic fluid with dissipation, with its energy-momentum tensor being
\begin{equation}\label{eq:imperfect fluid}
    \cT_{\mu\nu} = p g_{\mu\nu} + (\rho + p) u_\mu u_\nu + \Delta \cT_{\mu\nu}(\eta,\xi,\chi) \ ,
\end{equation}
where $g_{\mu\nu}$ is the spacetime metric, $\rho$, $p$, and $u_\mu$ denote the energy density, pressure, and velocity four-vector of the cosmic fluid, respectively.
The additional small term $\Delta \cT_{\mu\nu}$ stands for the deviation from perfect fluid caused by the dissipation, and can be generally characterized by three parameters, namely, the shear viscosity $\eta$, bulk viscosity $\xi$, and heat conduction $\chi$ \cite{Eckart:1940te,Weinberg:1971mx}. 
The presence of $\Delta \cT_{\mu\nu}$ modifies the equations of motion of the perturbations $\delta \rho$, $\delta p$, and $\delta u_\mu$, effectively introducing an additional decay rate proportional to $k^2$, which implies significant damping of small-scale perturbations \cite{Weinberg:1971mx}.

From microscopic perspectives, this dissipation originates from the diffusion and thermalization of particles in cosmic fluid .
During the RD era, the dominant source of dissipation is the shear viscosity, which is given by the specific particle interactions, i.e.,
\cite{Jeong:2014gna,Yu:2024xmz}
\begin{equation}\label{eq:shear viscosity}
   \eta(\tau)\simeq
   \frac{16}{45}\rho_\gamma t_\gamma 
   +\sum_{j=\nu,X,\, \ldots}\,\frac{4}{15}\rho_j t_j \,\Theta(\tau_{j,\mathrm{dec}}-\tau)+...\ .
\end{equation} 
Here, for each particle species $i$, $\rho_i$ is its energy density, $t_i$ is its mean free time, $\tau$ is the conformal time of the Universe, and $\tau_{i,\mathrm{dec}}$ is the decoupling time of the $i$-particles, defined by the condition $t_i(\tau_{i,\mathrm{dec}})\sim H^{-1}(\tau_{i,\mathrm{dec}})$ with $H$ being the cosmic expansion rate.
As demonstrated in \cref{eq:shear viscosity}, particles no longer contribute to the $\eta$ once they decouple, and the $\eta$ is primarily governed by the weakest-interacting particles (with the largest $t_i$) within the cosmic fluid.
Therefore, we can mainly consider the contributions of photons $\gamma$, neutrinos $\nu$ and some possible weakly-interacting particles \ac{BSM}, denoted by $X$, with their respective mean free times as follows
\begin{equation}\label{eq:cross section}
    t_\gamma=(n_{e^\pm} \langle \sigma_\mathrm{KN} v \rangle )^{-1}\ ,\ 
    t_\nu=(n_\nu \langle \sigma_\nu v \rangle)^{-1}\ ,\ 
    \mathrm{and}\ 
    t_X=(n_X \langle \sigma_X v \rangle)^{-1}\ .
\end{equation}
In \cref{eq:cross section}, $n_i$ is the number density and $\langle \sigma_i v \rangle$ is the thermally-averaged cross section for the corresponding interactions, i.e., $\langle \sigma_\mathrm{KN} v \rangle$ is the Klein-Nishina cross section for photon-electron scattering, $\langle \sigma_\nu v \rangle$ is the weak interaction cross section for neutrinos, and $\langle \sigma_X v \rangle$ is a model-dependent cross section given by the related new physics.

For the interest of \ac{GW} observations, we focus on the dissipative effects above $\sim \mathrm{keV}$.
In this case, the dissipation from photons is negligible \cite{Weinberg:1971mx}, and the $\eta$ is dominated by neutrinos and possible $X$-particles.
At the temperature $T$, the related quantities of particle species $j=\{\nu,X\}$ in \cref{eq:shear viscosity} are given by $\rho_j\sim g_j T^4$, $n_j \sim g_j T^3$ with $g_j$ the degrees of freedom, and $\langle \sigma_j v \rangle$ is generally parameterized at a reference energy scale $T_\ast$, i.e.,
\begin{equation}\label{eq:general cross section}
    \langle \sigma_j v \rangle
   = \langle \sigma_j v \rangle_{\ast}
   \left(\frac{T}{T_\ast}\right)^{m_j(T)}
   \ .
\end{equation}
As an example, we here consider neutrinos within the \ac{SM} and new $X$-particles interacting through the weak interaction mediated by some heavy gauge bosons $W'$.
Then \cref{eq:general cross section} can be specifically expressed as
\begin{align}\label{eq:specific cross section}
\langle \sigma_\nu v \rangle \simeq
\left\{
\begin{aligned}
& \ \alpha_F^2/T^2\ ,\ T\gg M_W \ ,
\\
& \ G_F^2 \, T^2\ ,\ T\lesssim M_W\ ,
\end{aligned}
\right.\ 
\mathrm{and}\ 
\langle \sigma_X v \rangle \simeq
\left\{
\begin{aligned}
& \ (\alpha')^2/T^2\ ,\ T\gg M_{W'} \ ,
\\
& \ (G')^2 \, T^2\ ,\ T\lesssim M_{W'}\ .
\end{aligned}
\right.\ 
\end{align}
Here, $M_W\sim100\, \mathrm{GeV}$ is the $W$ mass, $G_F\sim 10^{-5}\, \mathrm{GeV}^{-2}$ is the Fermi constant, and $G_F=\alpha_F/M_W^2$.
Similarly, $M_{W'}$ is the $W'$ mass with $M_{W'}> M_W$, and we have $G'=G_F(M_W/M_{W'})^2$ and $G'=\alpha'/M_{W'}^2$.
The different temperature dependence of $\langle \sigma_j v \rangle$ before and after the phase transitions stems from the fact that the corresponding mediated bosons are massless and massive, respectively.
With these particle interactions in the early Universe, we can determine the dissipation in cosmic fluid through \cref{eq:shear viscosity}.

\subsection{Evolution of cosmological scalar perturbations}\label{sec:2.2}

From now, let us move on to the effects of dissipation on cosmological perturbations.
In this paper, we adopt a perturbed Friedmann-Robertson-Walker metric in the conformal Newtonian gauge with the anisotropic stress being neglected, i.e.,
\begin{equation}\label{eq:metric}
    \ud s^2
    = a^2(\tau) \left\{
            - \left( 1 + 2 \phi \right) \ud \tau^2
            + \left[ (1 - 2 \phi) \delta_{ij} + \frac{1}{2} h_{ij} \right]
            \ud x^i \ud x^j
        \right\}\ ,
\end{equation}
where $a$ is the scale factor of the Universe, $\phi$ is the linear scalar perturbations, and $h_{ij}$ is the secondary transverse-traceless tensor perturbations sourced by $\phi$, known as \acp{IGW}.

Dissipation in cosmic fluid can significantly affect the evolution of $\phi$ inside the horizon.
We define the transfer function $\Phi(k,\tau)$ to describe the evolution of $\phi_\bk$, i.e.,
\begin{equation}\label{eq:T-def}
    \phi_\bk(\tau)
        = \frac{3+3w}{5+3w}\, \Phi(k,\tau) \, \zeta_\bk\ ,
\end{equation}
where $w=p/\rho$ is the parameter of the equation of state, with $w=1/3$ in the \ac{RD} era, $\zeta_\bk$ is primordial curvature perturbations, and $\bk$ denotes the Fourier mode of perturbations.
Given $\nabla^2 \phi \sim \delta \rho/\rho$, the damping of $\delta \rho$ in the presence of dissipation naturally leads to the damping of $\phi$, with their transfer function given by \cite{Jeong:2014gna}
\begin{equation}\label{eq:trans}
    \Phi(k,\tau)
    \simeq
    \frac{9}{(k \tau)^2}
    \left(
    \frac{\sqrt{3}}{k \tau}\sin{\frac{k \tau}{\sqrt{3}}}
    -\cos{\frac{k \tau}{\sqrt{3}}}
    \right)
    \,
    e^{-{k^2}/{k_D^2(\tau)}}\ ,
  \end{equation}
where $k_D$ is a typical damping scale.
For the scales $k\ll k_D$, \cref{eq:trans} goes back to the standard result in the \ac{RD} era.
While for the small scales $k\gg k_D$, the exponential damping factor greatly accelerates the decay of $\Phi(k,\tau)$, implying that dissipative effects dominates the evolution of $\phi$.
The parameter $k_D$ plays the central role in the following discussion.
In the early Universe, $k_D$ can be determined by the shear viscosity in cosmic fluid, i.e., \cite{Weinberg:1971mx}
\begin{equation}\label{eq:kD}
    k_D^{-2}\simeq 
    \int_0^\tau \, \mathrm{d} \bar{\tau}\ 
    \frac{2\eta}{3a\,(\rho+p)}
    \simeq \int_\infty^{T} \ud \bar{T}\, \left(-\frac{1}{aH\bar{T}}\right)\,  \frac{2\eta}{3a\,(\rho+p)}\ ,
\end{equation}
with $\eta$ given by \cref{eq:shear viscosity}. 
In the second step, we transform the integral variable from $\tau$ to $T$ by using the relation $H\sim T^2/M_\mathrm{pl}\sim 2/t$ and $\ud \tau/\ud T=(1/a)\, \ud t/\ud T \sim -1/(aHT)$, where $M_\mathrm{pl}$ is the reduced Planck mass and $t$ is the cosmic physical time.

\begin{figure*}[t]
    \centering
    \includegraphics[width=0.8\textwidth]{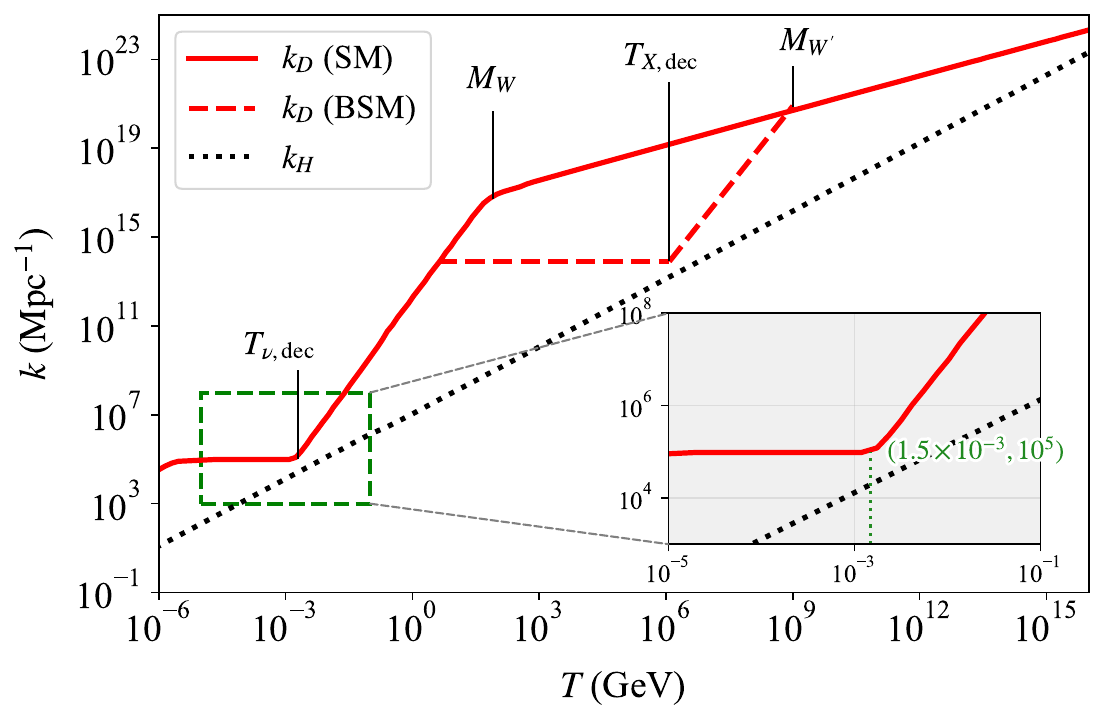}
    \caption{Temperature dependence of the damping scale $k_{D}$. 
    Red solid line stands for the \ac{SM} \cite{Jeong:2014gna}, while red dashed lines denote BSM. We plot the comoving Hubble horizon $k_\cH=aH$ for comparison.}
    \label{fig:kD_T}
\end{figure*}

In \cref{fig:kD_T}, we plot the temperature dependence of $k_D$ above $\sim \mathrm{keV}$.
Within the \ac{SM}, $k_D$ is dominated by neutrino interactions.
Using Eqs.~(\ref{eq:shear viscosity}-\ref{eq:specific cross section},\,\ref{eq:kD}) and neglecting the temperature dependence of the degrees of freedom in cosmic fluid, we have $k_D\propto T^{1/2}$ at $T\gtrsim M_W$, and $k_D\propto T^{5/2}$ at $T\lesssim M_W$.
After neutrinos decouple at $T\simeq 1.5\, \mathrm{MeV}$, $k_D$ remains roughly constant at $\sim 10^5\, \mathrm{Mpc}^{-1}$, until the photon interactions start to govern $k_D$ at $\sim \mathrm{keV}$.
Additionally, we also consider the case that $k_D$ is dominated by new $X$-particles with the interactions described by \cref{eq:specific cross section}.
It is straightforward to see that $k_D$ follows the temperature behavior similar to the neutrino-domination case but happens at higher energy scales.
In general, the damping of $\phi$ caused by any new particles can be analyzed by substituting their cross sections into Eqs.~(\ref{eq:shear viscosity},\,\ref{eq:kD}).

The evolution of $\phi_\bk$ inside the horizon depends on the ratio between $k_D$ and the horizon scale $k_\cH = \cH=aH$, where $\cH$ is the comoving Hubble parameter.
As illustrated in Fig.~\ref{fig:kD_T}, $k_D$ is usually much larger than $k_\cH$ but approaches $k_\cH$ near the decoupling time of weakly-interacting particles.
The latter can be briefly derived as follows.
Assuming that $\eta$ is dominated by the interaction of  $j$-particles given by \cref{eq:general cross section} with $T_\ast=T_{j,\mathrm{dec}}$ and $m_j$ a constant, $k_D$ at the decoupling of $j$-particles can be estimated as
\begin{equation}\label{eq:decouple}
    k_{D,j}^{-2}
    \, \simeq \, \int_\infty^{T_{j,\mathrm{dec}}} \ud T\, \left(-\frac{1}{aHT}\right)\, \frac{2\,\rho_j t_j}{15\, a\rho}\ 
    \simeq\,  \frac{1}{m_j+3}\times \frac{2\, \rho_j t_j}{ 15\, a^2H\rho}\, \bigg|_{T=T_{j,\mathrm{dec}}}
    \simeq\,  \frac{2\rho_j}{15(m_j+3)\rho}\, k^{-2}_{\cH,j}\ ,
\end{equation}
where we introduce $k_{D,j}=k_D(\tau_{j,\mathrm{dec}})$ and $k_{\cH,j}=k_\cH(\tau_{j,\mathrm{dec}})$ as short-hand notations.
To evaluate the integral above, we should notice that the integrand scales as $\sim T^{-(m_j+4)}$ and also apply the decoupling condition $t_j\sim H^{-1}$.
\cref{eq:decouple} tells us the important relation
\begin{equation}\label{eq:kDkH}
    \frac{k_{D,j}}{k_{\cH,j}}
    \simeq \sqrt{\frac{15(m_j+3)\rho}{2\rho_j}}\ .
\end{equation}
Note that though we neglect the change of degrees of freedom for simplicity, it is expected to modify the result only by an $\cO(1)$ factor.
\cref{eq:kDkH} generally implies that $k_{D,j}$ is comparable to $k_{\cH,j}$ as long as the abundance of $j$-particles is non-negligible at decoupling.
Therefore, the evolution of $\phi_\bk$ can be concluded as: the modes $k\sim k_{D,j}$ would undergo exponential decay soon after horizon reentry, while for other modes, the damping effects are negligible unless they are deep in the horizon.
This also aligns with the physical picture that particle interactions become weakest right before decoupling and cause the maximal dissipation.

\begin{figure*}[t]
    \centering
    \includegraphics[width=1\textwidth]{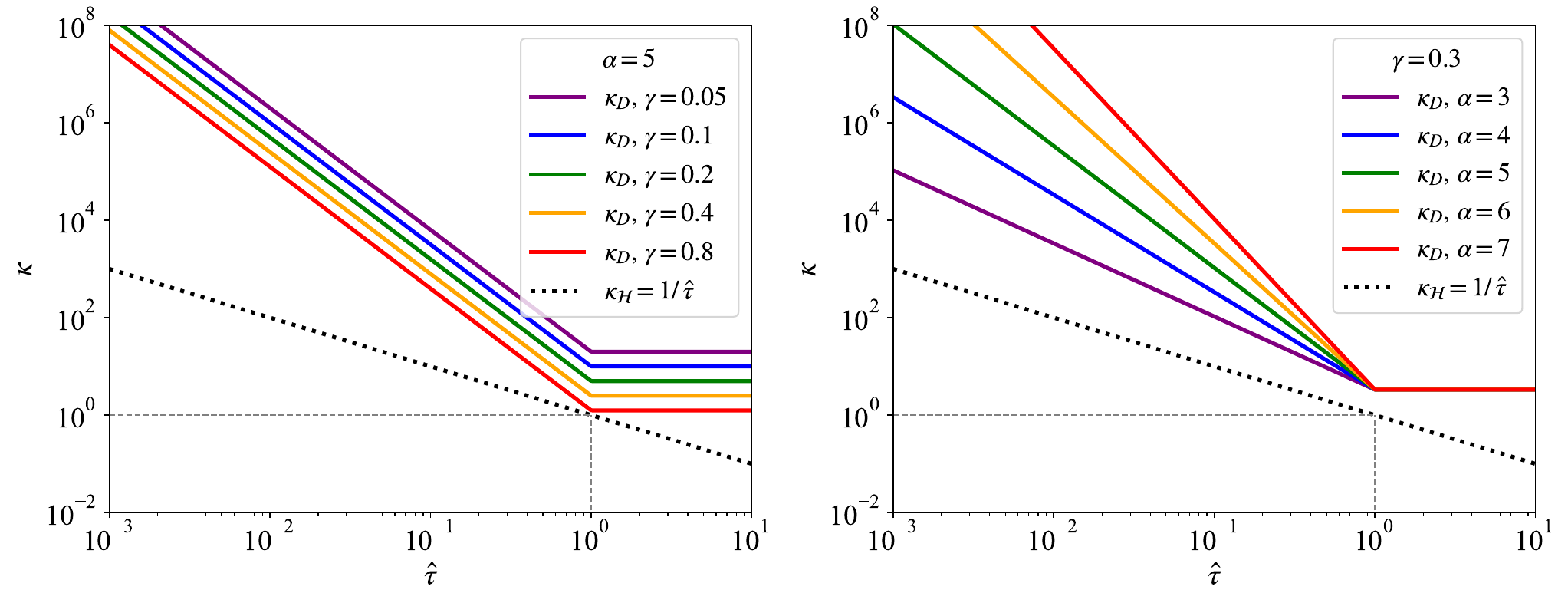}
    \caption{Damping scale $\kappa_D$ as a function of $\hat{\tau}$ near the particle decoupling for different values of $\alpha$ and $\gamma$. The comoving Hubble parameter $\kappa_\cH=1/\hat{\tau}$ is also plotted.}
    \label{fig:kD_tau}
\end{figure*}

In the subsequent discussion, instead of concentrating on specific particle species, we extract the main feature of $k_D$ and utilize a parameterization method to study dissipative effects. 
Here, we rewrite \cref{eq:trans} as the function of dimensionless variables $\kappa=k/k_{\cH,j}=k\tau_{j,\mathrm{dec}}$ and $\hat{\tau}=\tau/\tau_{j,\mathrm{dec}}$, i.e.,
\begin{equation}\label{eq:trans2}
    \Phi(\kappa,\hat{\tau})
    \simeq
    \frac{9}{(\kappa \hat{\tau})^2}
    \left(
    \frac{\sqrt{3}}{\kappa \hat{\tau}}\sin{\frac{\kappa \hat{\tau}}{\sqrt{3}}}
    -\cos{\frac{\kappa \hat{\tau}}{\sqrt{3}}}
    \right)
    \,
    e^{-{\kappa^2}/{\kappa_D^2(\hat{\tau})}}
    \ ,
\end{equation}
where the reduced damping scale is defined as $\kappa_D(\hat{\tau})=k_D(\tau)/k_{\cH,j}$.
Based on the discussion in last paragraph, we can focus on the damping effects near the decoupling time of $j$-particles, and $\kappa_D$ can be parameterized as
\begin{align}\label{eq:kappaD}
\kappa_D(\hat{\tau})
    \simeq
\left\{
\begin{aligned}
& \ {\gamma}^{-1}\, \hat{\tau}^{-\alpha/2} \ ,\ \ \hat{\tau}<1
\\
& \ {\gamma}^{-1}\ ,\ \ \hat{\tau}>1\ .
\end{aligned}
\right.\ 
\end{align}
The two dissipation parameters $\alpha$ and $\gamma$ in \cref{eq:kappaD} are defined as 
\begin{equation}\label{eq:alpha gamma}
    \hat{\tau}^{-\alpha/2}
    \equiv\frac{k_{D}(\tau)}{k_{D,j}}
    \simeq\hat{\tau}^{-(m_j+3)/2}\ ,\ 
    \mathrm{and}\     
    \gamma^{-1}
    \equiv\frac{k_{D,j}}{k_{\cH,j}}
    \simeq \sqrt{\frac{15(m_j+3)\rho}{2\rho_j}}\ ,
\end{equation}
respectively.
Fig.~\ref{fig:kD_tau} clearly shows how $\alpha$ and $\gamma$ determine the evolution of $\kappa_D$ near the $j$-particle decoupling. 
In what follows, we will use $\alpha$ and $\gamma$ as free parameters to characterize the properties of $\kappa_D(\hat{\tau})$, and they can be connected to the particle interaction and abundance through \cref{eq:alpha gamma}.

\subsection{Energy-density spectrum of induced gravitational waves}\label{sec:2.3}

The preparations in the previous two subsections allow us to further investigate dissipative effects in \acp{IGW}.
In \cref{eq:metric}, the \acp{IGW}, $h_{ij}$, can be transformed into Fourier space as
\begin{equation}
    h_{ij}(\tau,\bx)
    = \sum_{\lambda=+,\times}
        \int \frac{\ud^3 \bk}{(2\pi)^{3/2}}\, e^{i\bk\cdot\bx}\,
        \epsilon_{ij,\bk}^{\lambda}\, h_\bk^{\lambda}(\tau)\ ,
\end{equation}
where $\lambda=+,\times$ denotes two polarization modes of \acp{GW}, and the two polarization tensors $\epsilon_{ij,\bk}^{\lambda}$ are defined as $\epsilon^+_{ij,\bk}=[\epsilon_{i,\bk}\, \epsilon_{j,\bk}- \bar{\epsilon}_{i,\bk}\, \bar{\epsilon}_{j,\bk}]/\sqrt{2}$ and $\epsilon^\times_{ij,\bk}=[\epsilon_{i,\bk}\, \epsilon_{j,\bk}+ \bar{\epsilon}_{i,\bk}\, \bar{\epsilon}_{j,\bk}]/\sqrt{2}$, respectively, with $\epsilon_{i,\bk}$ and $\bar{\epsilon}_{i,\bk}$ forming an orthonormal basis perpendicular to $\bk$.
Based on cosmological perturbation theory, the motion of $h_\bk^{\lambda}$ satisfies
\begin{equation}\label{eq:IGW motion}
    h_\bk^{\lambda}{}\,''(\tau)
    + 2\cH\, h_\bk^{\lambda}{}\,'(\tau)
    +k^2 h_\bk^{\lambda}(\tau) 
    = 4\, \cS_\bk^{\lambda}(\tau)\ ,
\end{equation} 
where a prime stands for a derivative with respect to $\tau$, and the source term reads
\begin{equation}\label{eq:source}
    \cS_\bk^{\lambda}
    = \int \frac{\ud^3 \bq}{(2\pi)^{3/2}}\  \epsilon_{ij,\bk}^{\lambda}\, q_i q_j\times
        \left[
        2\phi_{\bk-\bq}\phi_\bq+
        (\cH^{-1} \phi'_{\bk-\bq}+ \phi_{\bk-\bq})
        (\cH^{-1} \phi'_{\bq}+ \phi_{\bq})
        \right]\ .
\end{equation}
By employing the Green’s function method \cite{Espinosa:2018eve,Kohri:2018awv}, \cref{eq:IGW motion} can be solved as
\begin{equation}\label{eq:ah}
    a(\tau)h_\bk^{\lambda}(\tau)
    =
    4\int^\tau \ud{\tau_1}\ 
    G_\bk(\tau,{\tau_1})\,
    a({\tau_1})\, \cS_\bk^{\lambda}({\tau_1})\ .
\end{equation}
Here, $G_\bk(\tau,{\tau_1})$ stands for the Green's function, which is defined as the solution of the Green's equation
\begin{equation}\label{eq:Green equation}
    G_\bk''(\tau,{\tau_1})+\left(k^2-\frac{a''(\tau)}{a(\tau)}\right)G_\bk(\tau,{\tau_1})
    =\delta(\tau-{\tau_1})\ ,
\end{equation}
and the expression of $G_\bk(\tau,{\tau_1})$ is obtained to be
\begin{equation}\label{eq:Green function}
    G_\bk(\tau,{\tau_1})
    =\sin{(k\tau-k{\tau_1})}/k\ .
\end{equation}
To relate \acp{IGW} with $\zeta$, we rewrite Eqs.~(\ref{eq:source},\,\ref{eq:ah}) in the form of
\begin{equation}\label{eq:h}
    h_\bk^{\lambda}(\tau)
    =4 \int \frac{\ud^3 \bq}{(2\pi)^{3/2}}\, Q^{\lambda}(\bk,\bq)\ 
    I(|\bk-\bq|,q,\tau)\, \zeta_{\bk-\bq}\, \zeta_\bq\ .
\end{equation}
Here, the projection factor $Q^{\lambda}(\bk,\bq)$ describes the geometric relations between the momenta and polarization tensors of the linear perturbations, being defined as
\begin{equation}
    Q^{\lambda}(\bk,\bq)=\epsilon_{ij,\bk}^{\lambda}\, q_i q_j/\, k^2
\end{equation}
and the kernel function $I(|\bk-\bq|,q,\tau)$ encodes the evolution information of \ac{GW} sources after inflation, which is given by
\begin{equation}\label{eq:I}
\begin{aligned}
    I(|\bk-\bq|,q,\tau)
    =\int_0^{k\tau}\ud(k{\tau_1})\, 
    \frac{a({\tau_1})}{a(\tau)}
    \, k G_\bk(\tau,{\tau_1})
    \, f(|\bk-\bq|,q,{\tau_1})\ .
\end{aligned}
\end{equation}
where $f(|\bk-\bq|, q, \tau)$ is constructed from the combination of transfer functions in \cref{eq:trans}, i.e.,
\begin{equation}\label{eq:f}
\begin{aligned}
   f(|\bk-\bq|,q,{\tau_1})
   &=\frac{4}{3}\, 
   \Phi(|\bk-\bq|,{\tau_1})\, \Phi(q,{\tau_1})
   +\frac{4}{9}\, 
   {\tau_1}^2\, 
   \Phi' (|\bk-\bq|,{\tau_1})\,
   \Phi' (q,{\tau_1})
   \\
   &\quad 
   +\frac{4}{9}\,
   \Big[
   {\tau_1} \, \Phi (|\bk-\bq|,{\tau_1})\,
   \Phi' (q,{\tau_1})
   +{\tau_1}\, \Phi' (|\bk-\bq|,{\tau_1})\,
   \Phi (q,{\tau_1})
   \Big]
   \ .
\end{aligned}
\end{equation} 
Here, we remind that $\Phi$ is dependent on dissipation parameters $\alpha$ and $\gamma$, and dissipative effects enter into \acp{IGW} through the exponential damping of \ac{GW} sources.

The \acp{IGW} naturally form a stochastic background in the early Universe. 
As the most basic observable to describe its statistical properties, the energy-density fraction spectrum of \acp{IGW}, ${\Omega}_\mathrm{gw}(k)$, defined as the \ac{GW} energy density $\rho_\mathrm{gw}$ per logarithmic wavenumber normalized by the Universe's critical energy density $\rho_\mathrm{c}$, can be calculated as follows
\begin{equation}\label{eq:energy-density}
    {\Omega}_\mathrm{gw}(k)
    =
    \frac{1}{\rho_c}\, 
    \frac{\ud \, \rho_\mathrm{gw}}{\ud\, \ln{k}}
    = 
    \frac{1}{48}
    \left(\frac{k}{\cH}\right)^2
    \sum_{\lambda=+,\times} 
    {\cP_h^{\lambda}(k)} \ ,
\end{equation}
where ${\cP_h^{\lambda}(k)}$ is the dimensionless power spectrum of \acp{IGW}, which is defined by the two-point correlation of $h^{\lambda}_{\bk}$, i.e.,
\begin{equation}\label{eq:Ph}
    \langle
        h^{\lambda}_{\bk}\,
        h^{\lambda'}_{\bk'}
    \rangle
    = \delta^{\lambda \lambda'}\,\delta^{(3)}(\bk + \bk')\,
        \frac{2\pi^2}{k^3}\, {\cP_h^{\lambda}(k)}\ .
\end{equation}
Eqs.~(\ref{eq:h},\,\ref{eq:energy-density},\,\ref{eq:Ph}) indicate $\Omega_\mathrm{gw}\sim \langle h^2 \rangle \sim \langle \cS^{\lambda}({\tau_1}) \cS^{\lambda}({\tau_2})  \rangle \propto \langle \zeta^4\rangle$.
With the assumption of the Gaussianity of $\zeta$, the four-point correlator $\langle \zeta^4\rangle$ factorizes into products of $\langle \zeta^2\rangle$ through Wick's theorem, expressed in terms of the primordial curvature power spectrum $\cP_\zeta$, which is defined by
\begin{equation}\label{eq:Pzeta}
    \langle
        \zeta_{\bk}\,
        \zeta_{\bk'}
    \rangle
    = \delta^{(3)}(\bk + \bk')\,
        \frac{2\pi^2}{k^3}\, {\cP_\zeta(k)}\ .
\end{equation}
For a parametric treatment of dissipative effects in \acp{IGW}, it is convenient to work with dimensionless variables, i.e., $u=|\bk-\bq|/k$, $v=q/k$, along with $\kappa$ and $\hat{\tau}$ introduced previously.
After straightforward calculations, $\Omega_\mathrm{gw}$ can be obtained as \cite{Espinosa:2018eve,Kohri:2018awv}
\begin{equation}\label{eq:ogw}
    \Omega_\mathrm{gw}(\kappa,\hat{\tau})
    =
    \frac{1}{6}
    \int _0^{\infty} \ud u
    \int _{\lvert 1-u \rvert} ^{ 1+u } \ud v
    \ Q^2(u,v)\
    \overbar{\cI^{2}(u,v,\kappa,\hat{\tau})} 
    \times \mathcal{P}_{\zeta}(u\kappa)
    \mathcal{P}_{\zeta}(v\kappa)\ .
\end{equation}
In \cref{eq:ogw}, the geometric factor $Q^2(u,v)$ is given by
\begin{equation}
    Q^2(u,v)= \delta^{\lambda\lambda'}Q^{\lambda}(\bk,\bq)\,Q^{\lambda'}(\bk,\bq)= \bigg[\frac{4 v^2-\left(1+v^2-u^2\right)^2}{4uv}\bigg]^2\ ,
\end{equation}
and $\cI(u,v,\kappa,\hat{\tau}) $ is defined by $(k/\cH)\, I(|\bk-\bq|,q,\tau)$, with the oscillation average of 
$\cI^{2}(u,v,\kappa,\hat{\tau})$ given by
\begin{equation}\label{eq:IIbar}
\begin{aligned}
    \overbar{\cI^2(u,v,\kappa,\hat{\tau})}
    =&
    \int^{\kappa\hat{\tau}}_0 \ud(\kappa\hat{\tau}_1)
    \int^{\kappa\hat{\tau}}_0 \ud(\kappa\hat{\tau}_2)\,
    \times\frac{1}{2}\,
    \bigg\{
    \prod_{i=1,2}
    \big[\kappa\hat{\tau}_i\, \cos{(\kappa\hat{\tau}_i)}\,f(u\kappa,v\kappa,\hat{\tau}_i)\big]
     \\
    &+
    \prod_{i=1,2}
    \big[\kappa\hat{\tau}_i\,\sin{(\kappa\hat{\tau}_i)}\,f(u\kappa,v\kappa,\hat{\tau}_i)\big]
    \bigg\}\  ,
\end{aligned}
\end{equation}
Here, $f(u\kappa,v\kappa,\hat{\tau}_i)$ is derived from \cref{eq:f} by substituting $|\bk-\bq|\rightarrow u\kappa$, $q\rightarrow v\kappa$ and $\tau_i\rightarrow\hat{\tau}_i$, and $\Phi(\kappa,\hat{\tau})$ is given by Eqs.~(\ref{eq:trans2},\,\ref{eq:kappaD}).

With Eqs.~(\ref{eq:ogw},\,\ref{eq:IIbar}), we can calculate the \ac{IGW} spectrum. 
In the dissipation-free case, i.e., $\kappa_D\rightarrow \infty$, $\cI(u,v,\kappa,\hat{\tau})$ has an analytical expression, namely \cite{Kohri:2018awv}
\begin{equation}\label{eq:cI}
\begin{aligned}
   &\cI(u,v,\kappa,\hat{\tau})
    =\frac{9}{u^3 v^3 (\kappa\hat{\tau})^3}
   \ \Bigg\{ 
   2 u v (\kappa\hat{\tau})^2 \cos{\frac{u\kappa\hat{\tau}}{\sqrt{3}}} \cos{\frac{v\kappa\hat{\tau}}{\sqrt{3}}} 
   -2 \sqrt{3} \,u (\kappa\hat{\tau}) \cos{\frac{u\kappa\hat{\tau}}{\sqrt{3}}} \sin{\frac{v\kappa\hat{\tau}}{\sqrt{3}}} 
   \\
   &\quad
   -2 \sqrt{3} \,v (\kappa\hat{\tau}) \sin{\frac{u\kappa\hat{\tau}}{\sqrt{3}}} \cos{\frac{v\kappa\hat{\tau}}{\sqrt{3}}} 
   +\Big[6+(u^2+v^2-3)(\kappa\hat{\tau})^2 \Big] \sin{\frac{u\kappa\hat{\tau}}{\sqrt{3}}} \sin{\frac{v\kappa\hat{\tau}}{\sqrt{3}}} 
   \Bigg\}
   \\
   &\quad
   +\frac{3(u^2+v^2-3)\sin{(\kappa\hat{\tau})}}{4u^3v^3}
   \Bigg\{
   (u^2+v^2-3)
   \Bigg[
   \mathrm{Ci} \left( \left( 1-\frac{v-u}{\sqrt{3}} \right) \kappa\hat{\tau} \right)
   +\mathrm{Ci} \left( \left( 1+\frac{v-u}{\sqrt{3}} \right) \kappa\hat{\tau} \right)
   \\
   &\quad
   -\mathrm{Ci} \left( \left| 1- \frac{v+u}{\sqrt{3}} \right| \kappa\hat{\tau} \right)
   -\mathrm{Ci} \left( \left( 1+\frac{v+u}{\sqrt{3}} \right) \kappa\hat{\tau} \right)
   +\ln \left(\left| \frac{3-(u+v)^2}{3-(u-v)^2}\right| \right)
   \Bigg]
   -4uv
   \Bigg\}
   \\
   &\quad
   -\frac{3(u^2+v^2-3)\cos{(\kappa\hat{\tau})}}{4 u^3 v^3}
   \ \Bigg\{ 
   (u^2+v^2-3)
   \Bigg[ 
   \mathrm{Si} \left( \left( 1-\frac{v-u}{\sqrt{3}} \right) \kappa\hat{\tau} \right)
   +\mathrm{Si} \left( \left( 1+\frac{v-u}{\sqrt{3}} \right) \kappa\hat{\tau} \right)
   \\
   &\quad
   -\mathrm{Si} \left( \left( 1-\frac{v+u}{\sqrt{3}} \right) \kappa\hat{\tau} \right)
   -\mathrm{Si} \left( \left( 1+\frac{v+u}{\sqrt{3}} \right) \kappa\hat{\tau} \right)
   \Bigg]
   \Bigg\}
   \ .
\end{aligned}
\end{equation} 
In \cref{eq:cI}, the sine/cosine-integral and logarithmic functions
come from the relations
\begin{equation}
    \int_0^x \ud \bar{x}\  \frac{\sin{A\bar{x}}}{\bar{x}}=\mathrm{Si}(Ax)\ ,\ \ 
    \mathrm{and}\ 
    \int_0^x \,\ud\bar{x} \ \frac{1-\cos{A\bar{x}}}{\bar{x}} = \gamma_E - \mathrm{Ci}(|Ax|)+\ln{|Ax|}\ ,
\end{equation}
where $\gamma_E\simeq 0.577$ is the Euler-Mascheroni constant. 
Taking the late-time limit, i.e.,  $\kappa\hat{\tau}\rightarrow\infty$, \cref{eq:cI} can be derived to the simple form
\begin{equation}
\begin{aligned}
    {\cI} 
    =
    \frac{3(u^2+v^2-3)^2}{4 u^3 v^3}
    \left[ \left(\ln{\left| \frac{3-(u+v)^2}{3-(u-v)^2} \right|} - \frac{4uv}{u^2+v^2-3}\right)\sin{(\kappa\hat{\tau})}-\pi \, \Theta(u+v-\sqrt{3})\,\cos{(\kappa\hat{\tau})}\right]\,,
\end{aligned}
\end{equation}
which avoids the highly-oscillating time integral in \cref{eq:IIbar} and largely simplifies the calculation of $\Omega_\mathrm{gw}(\kappa)$.
However, given $\kappa_D$ in \cref{eq:kappaD}, $\cI(u,v,\kappa,\hat{\tau})$ generally lacks any analytical expressions. 
We numerically calculate the $\Omega_\mathrm{gw}$ using an open Python package \texttt{vegas} \cite{Lepage:2020tgj}
\footnote{
In the numerical integral in \cref{eq:IIbar}, the choice of the upper limit $\kappa\hat{\tau}$ should balance the following two aspects.
On the one hand, $\kappa\hat{\tau}$ should be sufficiently large to ensure the late-time limit.
The production of \acp{IGW} is almost completed before this $\kappa\hat{\tau}$ and $\Omega_\mathrm{gw}(\kappa)$ remains constant with the further increase of $\kappa\hat{\tau}$.
On the other hand, for a too much large $\kappa\hat{\tau}$, the finite number of scattering points in the domain of integration may not accurately describe the highly oscillating integrand function, thus resulting in the distortion of numerical integral results.
}, as will be shown in \cref{sec:3}.

Before ending this section, we provide an intuitive physical picture about how dissipation affects the \ac{IGW} production.
Particularly, we introduce the conception of ``dissipation-induced lifetime'' of \ac{GW} sources, which is crucial for subsequent analysis. 
This conception is based on the following facts. 
Without dissipation, $\phi_\bk$ decays only slowly inside the horizon, roughly as $\sim 1/(\kappa\hat{\tau})^{2}$, remaining an active \ac{GW} source until $\kappa\hat{\tau}\gg 1$. 
However, with dissipation, the additional damping factor $\sim e^{-\kappa^2/\kappa_D^2(\hat{\tau})}$ can significantly accelerate the decay of $\phi_\bk$ for the mode $\kappa>\kappa_D(\hat{\tau})$. 
When only a rough analysis is conducted, we can simply assume that dissipation effects can be ignored before $\kappa\sim \kappa_D(\hat{\tau})$, while that \ac{GW} sources immediately vanish after $\kappa\sim \kappa_D(\hat{\tau})$. 
Mathematically, this is equivalent to making the following approximation in \cref{eq:trans2}
\begin{equation}\label{eq:tau_l}
    e^{-\kappa^2/\kappa_D^2(\hat{\tau})}\sim \Theta({\hat{\tau}}_l-{\hat{\tau}})\ ,\ \ 
    \mathrm{where}\ 
    \kappa\sim \kappa_D(\hat{\tau}_l)\ .
\end{equation}
Here, we refer $\hat{\tau}_l$ to as the ``dissipation-induced lifetime'' of $\phi_\bk$, which is determined by the mode $\kappa$ as well as the dissipation parameters $\alpha$ and $\gamma$. 
The \acp{IGW} could only be efficiently produced within the lifetimes of sources.
This approximation enables us to make use of the analytical results in \cref{eq:cI} and will largely simplify our analysis.

\section{Dissipative effects in the induced gravitational waves}\label{sec:3}

Based on the foundation established in \cref{sec:2}, we present the numerical results of $\Omega_\mathrm{gw}$ in the presence of dissipation for different types of $\cP_\zeta$ (i.e., scale-invariant, monochromatic, and log-normal ones). 
For each $\cP_\zeta$, we study the dependence of dissipative effects in $\Omega_\mathrm{gw}$ on the dissipation parameters $\alpha$ and $\gamma$.
Furthermore, we also reveal the physical origins behind these results, which are critical for us to link \ac{GW} observables to the early Universe. 
We summarize the basic properties of dissipative effects in \acp{IGW} at the end of this section.

\subsection{Scale-invariant curvature power spectrum}\label{sec:3.1}
Firstly, let us consider the simplest case where $\zeta$ has a scale-invariant spectrum
\begin{equation}\label{eq:Pzeta_SI}
    \cP_{\zeta}(\kappa)=\cA_{\zeta}\ , 
\end{equation}
where $\cA_\zeta$ is a spectral amplitude. 
This model can be an approximation of those flat spectra within the relevant range of scales. 
In the dissipation-free limit $\kappa_D\rightarrow\infty$, the corresponding \ac{IGW} spectrum is given by
\begin{equation}\label{eq:Ogw_SI}
    \Omega_\mathrm{gw}^{\kappa_D\rightarrow \infty}(\kappa)
    =0.822\,\cA^2_{\zeta}\ . 
\end{equation}
Since $\Omega_\mathrm{gw}^{\kappa_D\rightarrow \infty}$ is also scale-invariant, all the possible features in $\Omega_\mathrm{gw}$ could only come from dissipative effects, serving as an appropriate example to study them. 

\begin{figure*}[t]
    \centering
    \includegraphics[width=1\textwidth]{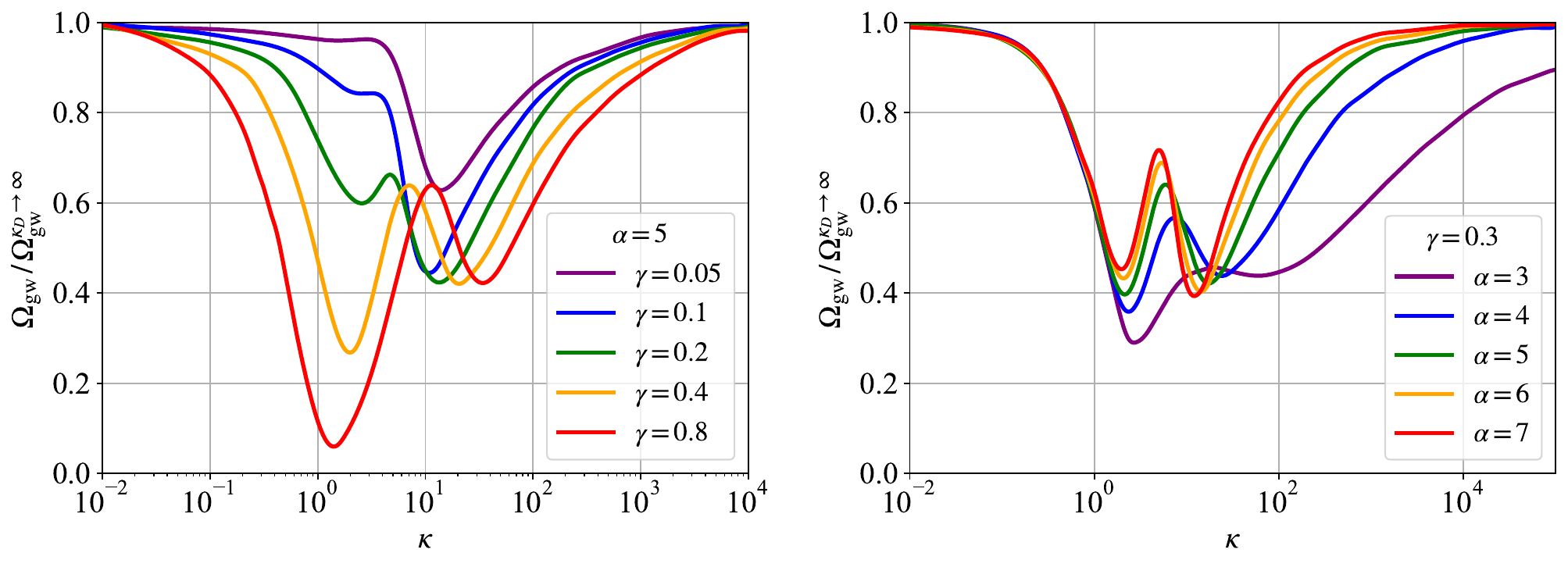}
    \caption{Energy-density spectrum $\Omega_\mathrm{gw}$ (normalized by $\Omega_\mathrm{gw}^{\kappa_D\rightarrow \infty}=0.822\,\cA_\zeta^2$) versus the scale $\kappa$ for scale-invariant primordial curvature power spectra. In the left panel, we fix $\alpha=5$ and take $\gamma=0.05, 0.1, 0.2, 0.4, 0.8$. In the right panel, we fix $\gamma=0.3$ and take $\alpha=3,4,5,6,7$. }
    \label{fig:SI}
\end{figure*}

In \cref{fig:SI}, we present the \ac{IGW} spectra with different dissipative parameters $\gamma$ and $\alpha$ in the left and right panel, respectively. 
As shown in both panels, dissipative effects generally introduce  multi(double) valleys, whose positions and shapes are dependent on $\gamma$ and $\alpha$. 
The left valley locates at $\kappa\sim \cO(1)$ with its position remaining relatively stable across the $\gamma$ and $\alpha$ range.
It becomes deeper as $\gamma$ increases and $\alpha$ decreases, and, particularly, can vanish for a sufficiently small $\gamma$.
The right valley locates at $\kappa\gtrsim \cO(10)$, and its position shows a more pronounced sensitivity, shifting towards larger $\kappa$ for increasing $\gamma$ and decreasing $\alpha$. 
Meanwhile, larger $\gamma$ and smaller $\alpha$ values tend to broaden the right valley, but its depth seems less sensitively dependent on these parameters.
For $\kappa\ll 1$ and $\kappa\rightarrow \infty$, the \ac{GW} spectrum goes back to the result without dissipation, implying that the dissipative effects become not important at the scales far away from the damping scale $k_{D,j}$.

\begin{figure*}[h]
    \centering
    \includegraphics[width=1\textwidth]{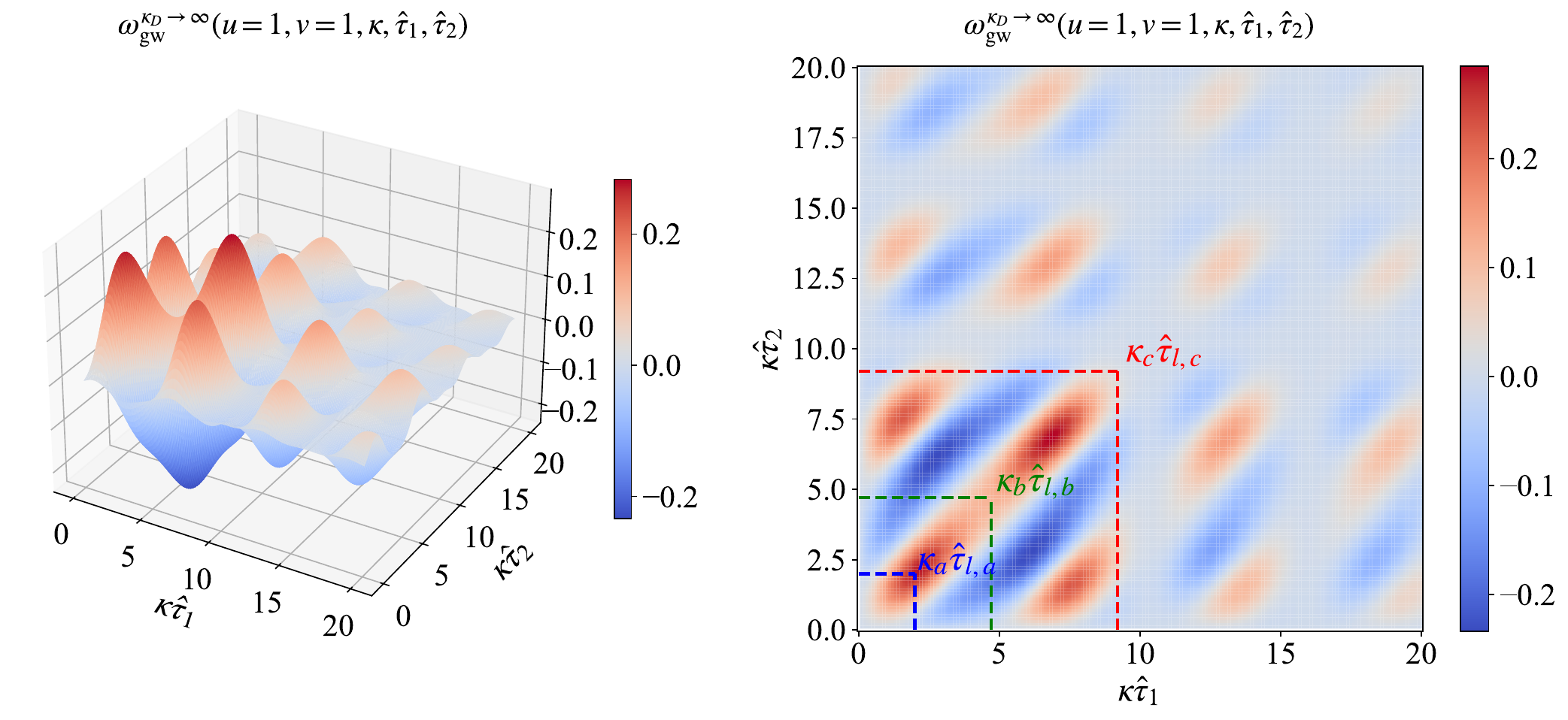}
    \caption{The 3D graph and heat map of $\omega_\mathrm{gw}^{\kappa_D\rightarrow\infty}(1,1,\kappa,\hat{\tau}_1,\hat{\tau}_2)$ as a function of $\kappa\hat{\tau}_1$ and $\kappa\hat{\tau}_2$. The red and blue regions correspond to the positive and negative values, respectively. In the right panel, we mark the integral regions given by the ``dissipation-induced lifetimes'' of the modes $\kappa_a=2$ (blue dashed lines), $\kappa_b=6$ (green dashed lines), and $\kappa_c=18$ (red dashed lines).}
    \label{fig:Physical_origin}
\end{figure*}

These features in $\Omega_\mathrm{gw}$ for a scale-invariant $\cP_\zeta$ arise from the oscillating properties of \ac{GW} sources during their dissipation-induced lifetimes. 
To better understand this, we recast Eqs.~(\ref{eq:ogw},\,\ref{eq:IIbar}) into the form
\begin{equation}\label{eq:wgw}
    \Omega_\mathrm{gw}(\kappa,\tau)
    =
    \int^{\kappa\hat{\tau}}_0 \ud(\kappa\hat{\tau}_1)
    \int^{\kappa\hat{\tau}}_0 \ud(\kappa\hat{\tau}_2)
    \left[
    \int _0^{\infty} \ud u
    \int _{\lvert 1-u \rvert} ^{ 1+u } \ud v
    \ \omega_\mathrm{gw}(u,v,\kappa,\hat{\tau}_1,\hat{\tau}_2)
    \right]\ ,
\end{equation}
where $[...]$ is the ``Green function'' contacting $\Omega_\mathrm{gw}$ to \ac{GW} sources at any given time $(\hat{\tau}_1,\hat{\tau}_2)$, with $\omega_\mathrm{gw}(u,v,\kappa,\hat{\tau}_1,\hat{\tau}_2)$ a combination of geometric factor, transfer functions, and primordial curvature power spectra, i.e.,
\begin{equation}\label{eq:wgw0}
\begin{aligned}
    \omega_\mathrm{gw}(u,v,\kappa,\hat{\tau}_1,\hat{\tau}_2)
    =&\,
    \frac{1}{12}\,
    Q^2(u,v)\,
    \bigg\{
    \prod_{i=1,2}
    \big[\kappa\hat{\tau}_i\, \cos{(\kappa\hat{\tau}_i)}\,f(u\kappa,v\kappa,\hat{\tau}_i)\big]
     \\
    &+
    \prod_{i=1,2} 
    \big[\kappa\hat{\tau}_i\,\sin{(\kappa\hat{\tau}_i)}\,f(u\kappa,v\kappa,\hat{\tau}_i)\big]
    \bigg\}\,
    \mathcal{P}_{\zeta}(u\kappa)
    \mathcal{P}_{\zeta}(v\kappa)\  .
\end{aligned}
\end{equation}
Based on the approximation in \cref{eq:tau_l}, within the lifetimes of \ac{GW} sources, \cref{eq:wgw0} can be treated the same as the dissipation-free case. 
Since $[...]$ in \cref{eq:wgw} is mainly contributed from $u,v\sim 1$ for scale-invariant $\cP_\zeta$, we consider the properties of $\omega_\mathrm{gw}^{\kappa_D\rightarrow\infty}(1,1,\kappa,\hat{\tau}_1,\hat{\tau}_2)$, which is plotted as a function of $\kappa{\tau}_1$ and $\kappa{\tau}_2$ in \cref{fig:Physical_origin}. 
Consisting of products of triangle functions, it oscillates in the $(\kappa\hat{\tau}_1,\kappa\hat{\tau}_2)$ plane, with red and blue regions representing the positive and negative contribution to \ac{IGW} spectrum, respectively. 
To obtain the $\Omega_\mathrm{gw}$ in the presence of dissipation, the time integral of $\Omega_\mathrm{gw}(\kappa)$ in \cref{eq:wgw} can be effectively limited in the region of $(0< \kappa\hat{\tau}_1<\kappa\hat{\tau}_l,\ 0< \kappa\hat{\tau}_2<\kappa\hat{\tau}_l)$. 
Due to the $\kappa$-dependence of $\tau_l$ in \cref{eq:tau_l}, this integral region varies with different scales.
Therefore, the oscillating phase of $\omega_\mathrm{gw}^{\kappa_D\rightarrow\infty}(1,1,\kappa,\hat{\tau}_1,\hat{\tau}_2)$ within the integral region could naturally induce peaks/valleys in $\Omega_\mathrm{gw}(\kappa)$. 
Meanwhile, since $\hat{\tau}_l$ also depends on $\alpha$ and $\gamma$, the locations and shapes of these peaks/valleys would change with these dissipation parameters.

\begin{figure*}[h]
    \centering
    \includegraphics[width=1\textwidth]{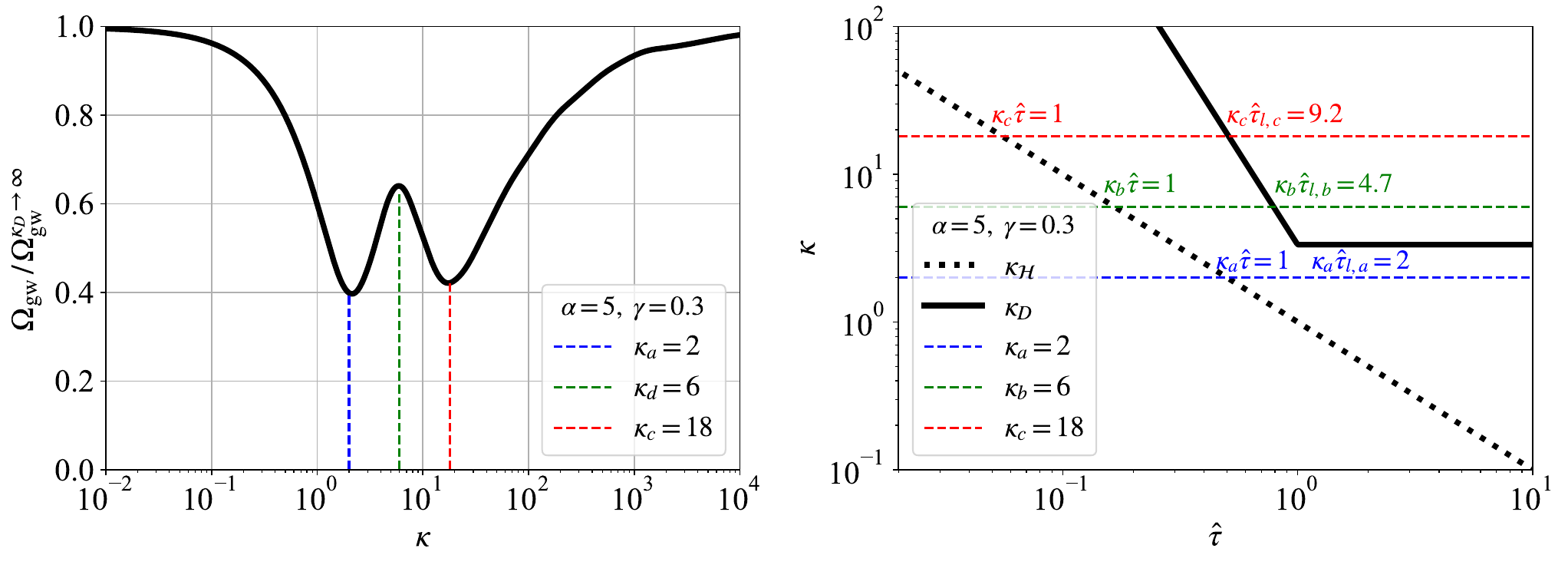}
    \caption{Left panel: Energy-density spectrum $\Omega_\mathrm{gw}$ (normalized by $\Omega_\mathrm{gw}^{\kappa_D\rightarrow \infty}=0.822\,\cA_\zeta^2$) versus the scale $\kappa$ for scale-invariant primordial curvature power spectra.
    Right panel: Damping scale $\kappa_D$ as a function of $\hat{\tau}$ near the particle decoupling. 
    In both panels, we mark the several reference scales (i.e., $\kappa_a=2$, $\kappa_b=6$, and $\kappa_c=18$), which correspond to the positions of peaks/valleys in $\Omega_\mathrm{gw}$.
    We set $\alpha=5$ and $\gamma=0.3$.}
    \label{fig:peaks}
\end{figure*}

Let us demonstrate in detail how the features in $\Omega_\mathrm{gw}$ are induced by dissipation in a specific example. 
In the left panel of \cref{fig:peaks}, we plot $\Omega_\mathrm{gw}(\kappa)$ with $\alpha=5$ and $\gamma=0.3$, and mark several reference scales (i.e., $\kappa_a=2$, $\kappa_b=6$, and $\kappa_c=18$), which roughly correspond to the locations of valleys/peaks in $\Omega_\mathrm{gw}$.  
According to \cref{eq:tau_l}, the dissipation-induced lifetimes of these modes are respectively given by $\kappa_a\hat{\tau}_{l,a}=2.0$, $\kappa_b\hat{\tau}_{l,b}=4.7$, and $\kappa_c\hat{\tau}_{l,c}=9.2$, as signed in the right panel of \cref{fig:peaks} \footnote{
In this example, $\kappa_D$ can approach $\kappa_a$ but $\kappa_D(\tau)=\kappa_a$ has no solution. 
Here, we simply take $\hat{\tau}_{l,a}$ as the initial moment when $\kappa_D$ is closest to $\kappa_a$, namely, $\hat{\tau}_{l,a}= 1$, as an approximation. 
This treatment will not affect our analysis.
}. 
Moreover, the corresponding integral regions $(0< \kappa\hat{\tau}_1<\kappa_i\hat{\tau}_{l,i},\ 0< \kappa\hat{\tau}_2<\kappa_i\hat{\tau}_{l,i})$ for $\kappa_i\ (i=a,b,c)$ are also marked in \cref{fig:Physical_origin}. 
Based on \cref{fig:Physical_origin}, the features in $\Omega_\mathrm{gw}(\kappa)$ can be derived from the oscillating behavior of $\omega_\mathrm{gw}^{\kappa_D\rightarrow\infty}$ in the integral regions.
(i) For $\kappa\ll1$ or $\kappa\rightarrow\infty$, the integral regions are given by $\kappa\hat{\tau}_l\rightarrow\infty$, leading to the same result as the dissipaton-free case.
(ii) Since the integral in $(0< \kappa\hat{\tau}_1<\kappa_a\hat{\tau}_{l,a},\ 0< \kappa\hat{\tau}_2<\kappa_a\hat{\tau}_{l,a})$ is less than that in dissipation-free case, there exhibits a valley in $\Omega_\mathrm{gw}(\kappa)$ at $\kappa\sim\kappa_a$. 
(iii) Since the integral region $(\kappa_a\hat{\tau}_{l,a}< \kappa\hat{\tau}_1<\kappa_b\hat{\tau}_{l,b},\ \kappa_a\hat{\tau}_{l,a}< \kappa\hat{\tau}_2<\kappa_b\hat{\tau}_{l,b})$ is dominated by a positive peak of $\omega_\mathrm{gw}^{\kappa_D\rightarrow\infty}$, the difference between $\Omega_\mathrm{gw}(\kappa_b)$ and $\Omega_\mathrm{gw}(\kappa_a)$ is positive. 
(iv) Since the integral region $(\kappa_b\hat{\tau}_{l,b}< \kappa\hat{\tau}_1<\kappa_c\hat{\tau}_{l,c},\ \kappa_b\hat{\tau}_{l,b}< \kappa\hat{\tau}_2<\kappa_c\hat{\tau}_{l,c})$ is dominated by two negative valleys of $\omega_\mathrm{gw}^{\kappa_D\rightarrow\infty}$, the difference between $\Omega_\mathrm{gw}(\kappa_c)$ and $\Omega_\mathrm{gw}(\kappa_b)$ is negative. 
(v) Similarly, we expect more peaks/valleys on $\Omega_\mathrm{gw}$ for $\kappa\gg\kappa_c$.
However, they are not that notable since the oscillations of $\omega_\mathrm{gw}^{\kappa_D\rightarrow\infty}$ in $(\kappa\hat{\tau}_1>\kappa_c\hat{\tau}_{l,c},\  \kappa\hat{\tau}_2>\kappa_c\hat{\tau}_{l,c})$ are much less pronounced than previous ones.

Similar analyses can be applied to general cases. 
Particularly, we find that even the following simple estimate can decently describe the features on $\Omega_\mathrm{gw}(\kappa)$. 
For $\kappa\hat{\tau}_1, \kappa\hat{\tau}_2 \gg 1$, the oscillating behavior of $\omega_\mathrm{gw}^{\kappa_D\rightarrow\infty}(1,1,\kappa,\hat{\tau}_1,\hat{\tau}_2)$ can be roughly described as
\begin{equation}\label{eq:wgw2}
    \omega_\mathrm{gw}^{\kappa_D\rightarrow\infty}(1,1,\kappa,\hat{\tau}_1,\hat{\tau}_2)
    \sim 
    \frac{1-\cos{\left({2\kappa\hat{\tau}_1}/{\sqrt{3}}\right)}}{\kappa\hat{\tau}_1}
    \times
    \frac{1-\cos{\left({2\kappa\hat{\tau}_2}/{\sqrt{3}}\right)}}{\kappa\hat{\tau}_2}
\end{equation}
If $\kappa$ is the location of peaks/valleys, the corresponding integral region $(0< \kappa\hat{\tau}_1<\kappa\hat{\tau}_{l},\ 0< \kappa\hat{\tau}_2<\kappa\hat{\tau}_{l})$ should roughly end at the zero point of \cref{eq:wgw2}, i.e., 
\begin{equation}\label{eq:zero}
    \kappa\hat{\tau}_l(\kappa,\alpha,\gamma)\simeq \sqrt{3}\, n\pi\ ,\ n=1,2...\ .
\end{equation}
Here, the dissipation-induced lifetime for mode $\kappa$ can be obtained by substituting $\kappa_D={\gamma}^{-1}\, \hat{\tau}^{-\alpha/2}$ into \cref{eq:tau_l}, namely
\begin{equation}\label{eq:taul}
    \hat{\tau}_l(\kappa,\alpha,\gamma)=(\gamma\kappa)^{-2/\alpha}\ .
\end{equation}
Combining Eqs.~(\ref{eq:taul},\,\ref{eq:zero}), the locations of peaks/valleys are given by
\begin{equation}\label{eq:peaks}
    \kappa\simeq(\sqrt{3}\, n\pi)^{\alpha/(\alpha-2)}\gamma^{2/(\alpha-2)}\ ,
    \ n=1,2...\ .
\end{equation}
In \cref{table:peaks}, we list the locations of the peaks (for $n=1$) and valleys (for $n=2$) that predicted by \cref{eq:peaks}, with the same values of $\alpha$ and $\gamma$ as those in \cref{fig:SI}. 
It can be verified that the results given by \cref{eq:peaks} approximately consist with the numerical results in \cref{fig:SI}.
This implies that our estimate captures the main properties of dissipative effects in \acp{IGW}. 

\begin{table}[h]
\centering
\begin{tabular}{|c|c|c|c|c|c|c|}
     \hline
     \multirow{3}*{$\alpha=5$}  &  
        & 
     $\gamma=0.05$ &
     $\gamma=0.1$ &
     $\gamma=0.2$ &
     $\gamma=0.4$ &
     $\gamma=0.8$ 
     \\
    \cline{2-7}
     ~  & $n=1$ (peak)  & 
     2.3 & 
     3.6 &
     5.8 &
     9.1 &
     14.5
     \\
    \cline{2-7}
     ~  & $n=2$ (valley) & 
     7.3 & 
     11.5 &
     18.2 &
     29.0 &
     46.1
     \\
    \hline\hline
     \multirow{3}*{$\gamma=0.3$}  &   
      & 
     $\alpha=3$ &
     $\alpha=4$ &
     $\alpha=5$ &
     $\alpha=6$ &
     $\alpha=7$ 
     \\
    \cline{2-7}
     ~  & $n=1$ (peak)  & 
     14.5 & 
     8.9 &
     7.5 &
     7.0 &
     6.6
     \\
    \cline{2-7}
     ~  & $n=2$ (valley) & 
     116.0 & 
     35.5 &
     24.0 &
     19.7 &
     17.5
     \\
    \hline
\end{tabular}
\caption{Positions of peaks/valleys on $\Omega_\mathrm{gw}(\kappa)$ for different dissipation parameters $\alpha$ and $\gamma$, as predicted by \cref{eq:peaks}.}\label{table:peaks}
\end{table}

\subsection{Monochromatic curvature power spectrum}\label{sec:3.2}
We then consider another case that $\zeta$ has a monochromatic spectrum, which can be modeled by the Dirac function, i.e.,
\begin{equation}\label{eq:Pzeta_D}
    \cP_{\zeta}(\kappa)
    =\cA_{\zeta}\ \delta\left(\ln{\frac{\kappa}{\kappa_\zeta}}\right)\ , 
\end{equation}
where $\cA_\zeta$ is an overall normalized factor and $\kappa_\zeta$ gives the peak wavenumber. 
This serves as an idealized approximation for a sharply-peaked spectrum. 
In the dissipation-free limit, the \ac{IGW} spectrum is given by
\begin{equation}\label{eq:Ogw_D}
\begin{aligned}
    \Omega_\mathrm{gw}^{\kappa_D\rightarrow \infty}(\kappa)
    =&\ \frac{3\cA^2_{\zeta}}{1024} \,
    \Theta\left( 2-\tilde{\kappa}\right)\,
    \tilde{\kappa}^2\,
     \left(4-\tilde{\kappa}^2\right)^2
     \left(3\tilde{\kappa}^2-2\right)^2
     \\
     &
     \ \times
     \left\{\pi^2\,
     \left(3\tilde{\kappa}^2-2\right)^2
     \Theta\left( 2\sqrt{3}-3\tilde{\kappa} \right)
     +
     \left[
     4+\left(3\tilde{\kappa}^2-2\right)\,
     \ln{\left|\frac{4}{3\tilde{\kappa}^2}-1\right|}
     \right]^2
     \right\}\ ,
\end{aligned}
\end{equation}
where $\tilde{\kappa}={\kappa}/{\kappa_\zeta}$.
This spectrum is well-known as the following properties: 
(i) a sharp peak at the logarithmic singularity $\kappa=(2/\sqrt{3})\,\kappa_\zeta$; 
(ii) a sharp valley near the zero point $\kappa=\sqrt{2/3}\,\kappa_\zeta$; 
(iii) a logarithmic running $\sim \kappa^2\ln^2{\kappa}$ in the infrared region.

\begin{figure*}[t]
    \centering
    \includegraphics[width=1\textwidth]{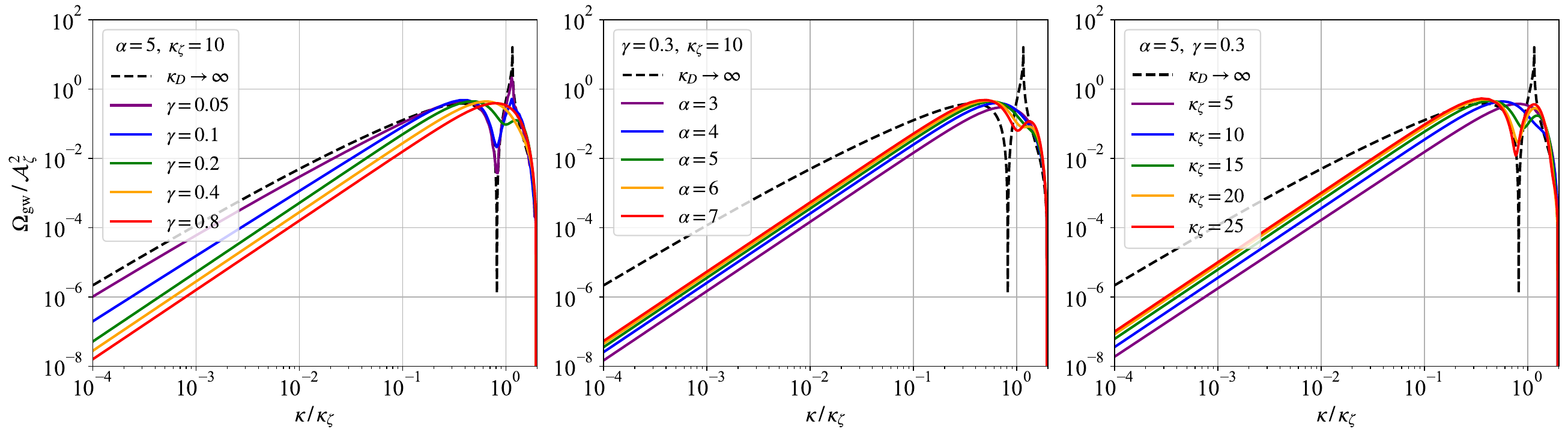}
    \caption{Energy-density spectrum $\Omega_\mathrm{gw}$ normalized by $\cA_\zeta^2$ versus the scale $\kappa/\,\kappa_\zeta$ for monochromatic primordial curvature power spectra. In the left panel, we fix $\alpha=5,\, \kappa_\zeta=10$ and take $\gamma=0.05,0.1,0.2,0.4,0.8$. In the middle panel, we fix $\gamma=0.3,\, \kappa_\zeta=10$ and take $\alpha=3,4,5,6,7$. In the right panel, we fix $\alpha=5,\,\gamma=0.3$ and take $\kappa_\zeta=5,10,15,20,25$. }
    \label{fig:Delta}
\end{figure*}

\cref{fig:Delta} displays the \ac{IGW} spectra for a monochromatic $\cP_\zeta$ with different choices of dissipation parameters $\alpha$, $\gamma$, and the peak position $\kappa_\zeta$. 
Each panel varies one parameter while keeping the other two fixed.
Across all panels in our plots, dissipative effects induces several common modifications compared to the spectrum given by \cref{eq:Ogw_D}:
(i) The resonant peak at $\kappa = ({2}/{\sqrt{3}}) \,\kappa_\zeta$ is suppressed and broadened.
This suppression becomes more pronounced for larger $\gamma$ and smaller $\alpha$, $\kappa_\zeta$.
(ii) The valley near $\kappa = \sqrt{{2}/{3}} \,\kappa_\zeta$ is progressively smoothed out.
The effect is enhanced by increasing $\gamma$ and descreasing $\alpha$, $\kappa_\zeta$.
This valley structure could even disappear under sufficiently strong dissipation.
(iii) The infrared region generally transitions from a logarithmic running $\sim \kappa^2 \ln^2{\kappa}$ to a steeper scaling $\sim \kappa^2$, indicating stronger suppression of long-wavelength modes due to the dissipation.
The onset of the $\kappa^2$ infrared scaling shifts to larger $\kappa/\kappa_\zeta$ as $\gamma$ increases and $\alpha$, $\kappa_\zeta$ decrease. 

These properties on $\Omega_\mathrm{gw}$ for monochromatic $\cP_\zeta$ have the same physical origin as those for scale-invariant $\cP_\zeta$, that is, the shorten of the lifetime of \ac{GW} sources. 
Below, we analyze respectively how dissipative effects modifies the above three properties of $\Omega_\mathrm{gw}$ for monochromatic $\cP_\zeta$. 
Here, we still consider the approximation in \cref{eq:tau_l}, which enables us to make the most use of the analytical expression in \cref{eq:cI} with $u=v=\kappa_\zeta/\kappa$. 
Thus, compared to Ref.~\cite{Domenech:2025bvr}, we provide a more intuitive analysis and do not require the power-law scaling of $\kappa_D(\hat{\tau})$. 

(i) The sharp peak at $\kappa = (2/{\sqrt{3}})\,\kappa_\zeta$ arises from the resonant amplification of scalar perturbations, where $1/\sqrt{3}$ is the sound speed and the factor 2 comes from the second-order effect. 
To carefully discuss the divergence of this peak, we examine the behavior of \cref{eq:cI} near the peak, i.e., $u=v\rightarrow{\sqrt{3}}/2$, and find it is dominated by the following terms
\begin{equation}\label{eq:I_delta}
    \cI(u,v,\kappa,\hat{\tau})
    \sim
    -\mathrm{Ci}\left(\left|1-\frac{2u}{\sqrt{3}}\right|\kappa\hat{\tau}\right)
    +\ln{\left(\left|1-\frac{2u}{\sqrt{3}}\right|\right)}\ .
\end{equation}
In the dissipation-free case, we usually consider the limit $\kappa\hat{\tau}\rightarrow\infty$ in \cref{eq:I_delta}, where the cosine-integral term vanishes while the leaving logarithmic singularity results in a divergent peak \footnote{This divergence may be kind of subtle, because $\left|1-\frac{2u}{\sqrt{3}}\right|\kappa\hat{\tau}$ here is actually a ``$0\cdot\infty$''-type indeterminate form. Anyway, this divergent peak usually does not cause serious disasters, since it is integrable and will disappears when replacing the unphysical Dirac-function $\cP_\zeta$ by a realistic one with finite spectral width. } \cite{Kohri:2018awv}.
However, the situation is differenet in the presence of dissipation.
Considering the finite lifetimes of \ac{GW} sources, the cosine-integral term cannot be neglected. 
Particularly, in the limits of $u=v\rightarrow{\sqrt{3}}/2$ and $\hat{\tau}\rightarrow\hat{\tau}_l(\kappa_\zeta,\alpha,\gamma)$, this cosine-integral term will tend to infinity and cancel out the logarithmic divergence term, ultimately giving a finite result
\begin{equation}\label{eq:I_delta_dis}
    \cI[u=v\rightarrow{\sqrt{3}}/2,\kappa,\hat{\tau}\rightarrow\hat{\tau}_l(\kappa_\zeta,\alpha,\gamma)]
    \sim
    -\gamma_E
    -\ln{\left[\kappa\hat{\tau}_l(\kappa_\zeta,\alpha,\gamma)\right]}\ .
\end{equation}
Here, we have used the relation
\begin{equation}\label{eq:lim}
    \lim_{x\rightarrow 0}\,[\,\mathrm{Ci}(x)-\ln{x}\,]=\gamma_E\ .
\end{equation}
According to \cref{eq:I_delta_dis}, the peak at $\kappa = (2/{\sqrt{3}})\,\kappa_\zeta$ is always finite even for a Dirac-function $\cP_\zeta$. 
Moreover, the suppression of this peak is more significant for larger dissipation (with smaller $\kappa\hat{\tau}_l(\kappa_\zeta,\alpha,\gamma)$ in \cref{eq:I_delta_dis}). 

(ii) Different from (i), the sharp valley at $\kappa=\sqrt{2/3}\,\kappa_\zeta$ is due to resonant cancellation of scalar perturbations. 
With dissipation, the reason of the disappearance of this valley is easy to point out: 
In \cref{eq:cI}, $u=v=\sqrt{3/2}$ is the solution of $\cI(u,v,\kappa,\hat{\tau})=0$ only for $\kappa\hat{\tau}\rightarrow\infty$. 
If the lifetimes of \ac{GW} sources are not sufficient large, the $\cO(1/(\kappa\hat{\tau}))$ terms in \cref{eq:cI} cannot be neglected, making $\kappa=\sqrt{2/3}\,\kappa_\zeta$ no longer a zero point of $\Omega_\mathrm{gw}$.

(iii) The infrared behavior of \ac{IGW} spectrum is dependent on $\cI(u,v,\kappa,\hat{\tau})$ in the limit  $u=v=\kappa_\zeta/\kappa\gg 1$, whose leading terms are given by
\begin{equation}\label{eq:IIR}
    {\cI(u,v,\kappa,\hat{\tau})}
    \sim 
    \frac{1}{u^2}
    \left[
    \mathrm{Ci}\left(\kappa\hat{\tau}\right)
    +\ln{\left(\frac{2u}{\sqrt{3}}\right)}\right]\ ,
\end{equation}
Without dissipation, we can ignore the cosine-integral terms in the limit $\kappa\hat{\tau}\rightarrow\infty$, leading to ${\cI}\sim u^{-2}\ln{u}$. 
Therefore, the \ac{IGW} spectrum in the infrared region behaves as $\Omega_\mathrm{gw}\sim (u^{-2}\ln{u})^2\,|_{u=\kappa_\zeta/\kappa} \times ({\kappa_\zeta}/{\kappa})^2 \sim \kappa^2\ln^2{\kappa}$. 
In the presence of dissipation, similar to (i), the modification of the infrared behavior also comes from the important contribution of cosine-integral terms in \cref{eq:IIR}. 
Taking $\hat{\tau}\rightarrow\hat{\tau}_l(\kappa_\zeta,\alpha,\gamma)$, \cref{eq:IIR} is roughly given by
\begin{equation}\label{eq:IIRdis}
\begin{aligned}
    &{\cI[u=v\gg1,\kappa\gg\kappa_\zeta,\hat{\tau}\rightarrow\hat{\tau}_l(\kappa_\zeta,\alpha,\gamma)]}
    \\
    &\sim 
    \frac{1}{u^2}
    \left\{
    \mathrm{Ci}\left[\frac{\kappa}{\kappa_\zeta}\kappa_\zeta\hat{\tau}_l(\kappa_\zeta,\alpha,\gamma)\right]
    -\ln{\left(\frac{\sqrt{3}\,\kappa}{2\,\kappa_\zeta}\right)}
    \right\}\ 
    \sim \ 
    \frac{1}{u^2}
    \left\{
    \gamma_E
    +\ln{\left[\frac{2\,\kappa_\zeta\hat{\tau}_l(\kappa_\zeta,\alpha,\gamma)}{\sqrt{3}}\right]}
    \right\}\ ,
\end{aligned}
\end{equation}
where we have used \cref{eq:lim} in the last step. 
Considering that $\kappa_\zeta\hat{\tau}_l(\kappa_\zeta,\alpha,\gamma)$ is not dependent on $\kappa$, \cref{eq:IIRdis} gives ${\cI}\sim u^{-2}$ without the logarithmic running, and the \ac{IGW} spectrum behaves as $\Omega_\mathrm{gw}\sim (u^{-2})^2\,|_{u=\kappa_\zeta/\kappa} \times ({\kappa_\zeta}/{\kappa})^2 \propto \kappa^2$ for $\kappa/\kappa_\zeta\rightarrow 0$. 
Furthermore, from \cref{eq:IIRdis}, it is easy to find the transition point from $\sim\kappa^2\ln^2{\kappa}$ scaling to $\sim\kappa^2$ scaling is determined by the following parameter
\begin{equation}
    \kappa_\zeta\hat{\tau}_l(\kappa_\zeta,\alpha,\gamma)
    =\gamma^{-2/\alpha}\kappa_\zeta^{1-2/\alpha}.
\end{equation}
Given smaller $\kappa_\zeta\hat{\tau}_l(\kappa_\zeta,\alpha,\gamma)$,  \cref{eq:IIRdis} could hold for larger $\kappa/\kappa_\zeta$, so the onset of the $\kappa^2$ scaling shifts to larger $\kappa/\kappa_\zeta$ for increasing $\gamma$ and decreasing $\alpha$, $\kappa_\zeta$. 
These properties are consistent with the numerical results in \cref{fig:Delta}.

\subsection{Log-normal curvature power spectrum}\label{sec:3.3}

In this subsection, we consider that $\zeta$ has a log-normal primordial curvature power spectrum
\begin{equation}\label{eq:Pzeta_LN}
    \cP_{\zeta}(\kappa)
    =\frac{\cA_{\zeta}}{\sqrt{2\pi}\sigma}\  e^{-\frac{1}{2 \sigma^2}\ln^{2} (\kappa/\kappa_\zeta)}\ , 
\end{equation}
where $\kappa_{\zeta}$ gives the peak wavenumber, $\cA_\zeta$ is the spectral amplitude at $\kappa_{\zeta}$, and $\sigma$ represents the spectral width. 
It could model a peaked spectrum with a finite width, and can be considered as a compromise between the previous two extreme cases.
In the dissipation-free case, the \ac{IGW} spectrum also presents as a wide peak with $\Omega_\mathrm{gw}^{\kappa_D\rightarrow \infty}\sim \cA_\zeta^2/(2\pi\sigma^2)$ at $\kappa\sim \kappa_\zeta$ and  $\Omega_\mathrm{gw}^{\kappa_D\rightarrow \infty}\sim \kappa^3\ln^2{\kappa}$ in the infrared region. 

\begin{figure*}[t]
    \centering
    \includegraphics[width=0.96\textwidth]{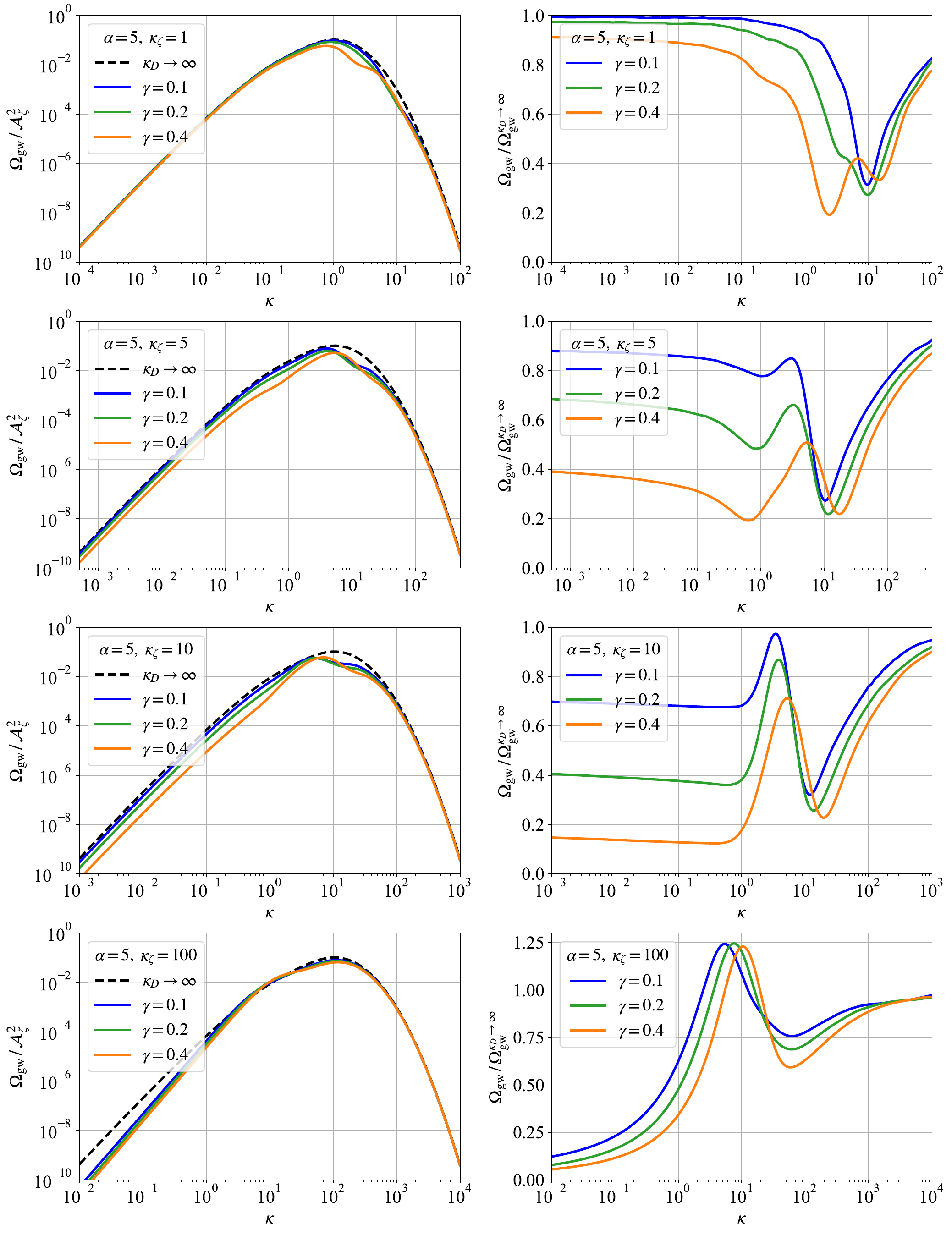}
    \caption{Energy-density spectrum $\Omega_\mathrm{gw}$ normalized by $\cA_\zeta^2$ (left column) and normalized by $\Omega_\mathrm{gw}^{\kappa_D\rightarrow \infty}$) (right column) versus the scale $\kappa$ for log-normal primordial curvature power spectra. 
    In both panels, we fix $\alpha=5$ and take $\gamma=0.1,0.2,0.4$. From the first to the fourth row, we respectively take $\kappa_\zeta=1,5,10,100$. }
    \label{fig:LN_gamma_kappac}
\end{figure*}

\begin{figure*}[t]
    \centering
    \includegraphics[width=0.96\textwidth]{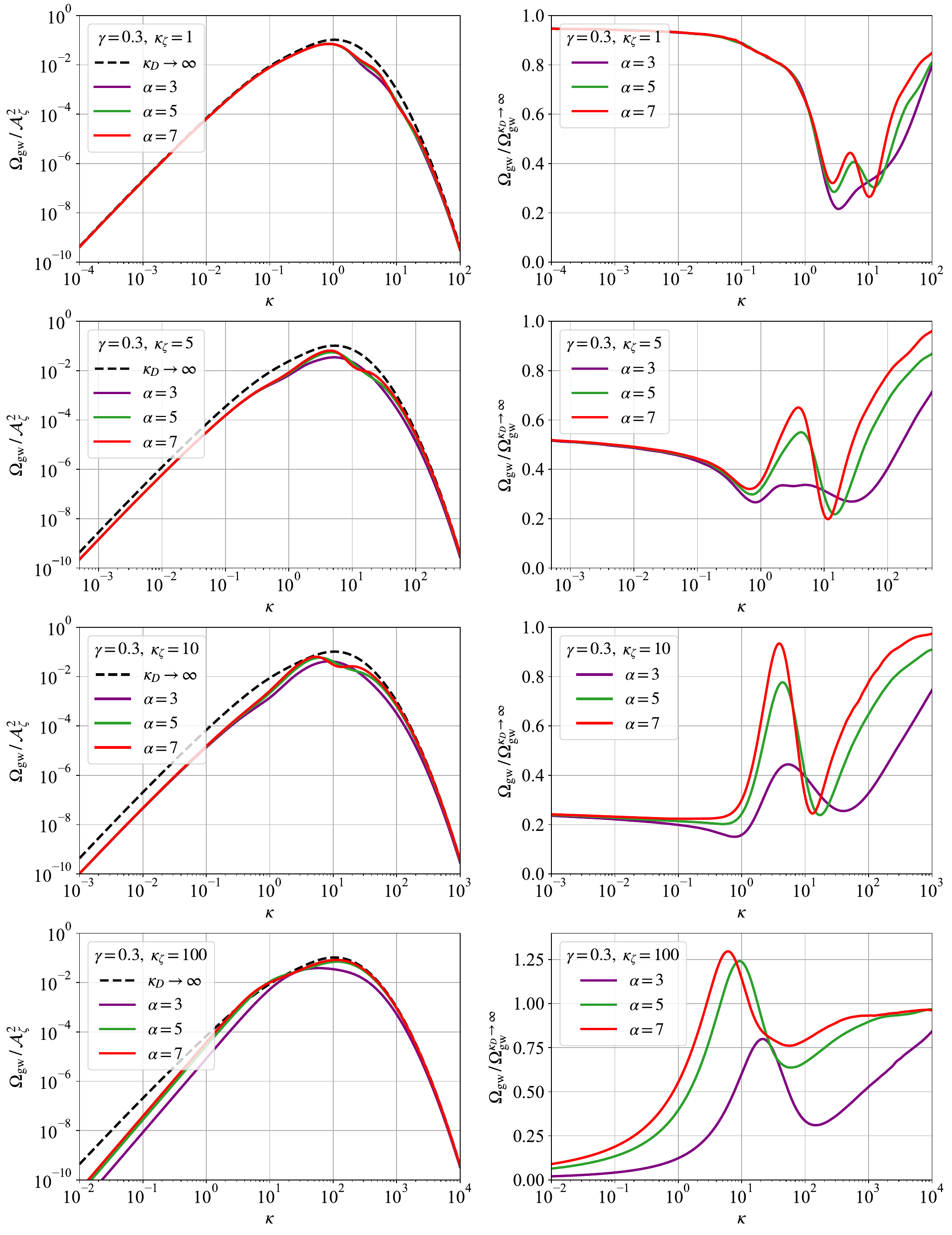}
    \caption{Energy-density spectrum $\Omega_\mathrm{gw}$ normalized by $\cA_\zeta^2$ (left column) and normalized by $\Omega_\mathrm{gw}^{\kappa_D\rightarrow \infty}$) (right column) versus the scale $\kappa$ for log-normal primordial curvature power spectra. 
    In both panels, we fix $\gamma=0.3$ and take $\alpha=3,5,7$. From the first to the fourth row, we respectively take $\kappa_\zeta=1,5,10,100$. }
    \label{fig:LN_alpha_kappac}
\end{figure*}

\cref{fig:LN_gamma_kappac} and \cref{fig:LN_alpha_kappac} show dissipative effects in \acp{IGW} for a log-normal $\cP_\zeta$, focusing on the roles of the dissipation parameters $\gamma$ and $\alpha$, respectively. 
In each figure, the left and right columns respectively depict the \ac{IGW} spectra normalized by $\mathcal{A}_\zeta^2$ and $\Omega_\mathrm{gw}^{\kappa_D\rightarrow \infty}$, and the four rows from top to bottom depict the cases where the peaks of $\cP_\zeta$ locate at $\kappa_\zeta = 1, 5, 10, 100$. 
We set $\sigma=1$ for all above cases. 
As shown in \cref{fig:LN_gamma_kappac} and \cref{fig:LN_alpha_kappac}, dissipative effects in this case exhibit combined properties of those in \cref{sec:3.1} and can also be modified by the shape of $\cP_\zeta$ itself. 
We analysis them as follows. 

(i) If dissipative effects happen in the ``peak region'' (i.e.,  $\kappa_\zeta\sim 1$), they are mainly manifested as amplitude suppression in $\kappa\gtrsim 1$, while the change in the infrared region is relatively small \footnote{
The properties of dissipative effects for $\kappa_\zeta\ll1$ are similar to those in the ``peak region''. We do not include $\kappa_\zeta\ll 1$ in our plots since $\Omega_\mathrm{gw}(\kappa\gtrsim 1)$ is too small in this case. 
}. 
Particularly, they generally lead to a similar peak/valley structure with the case of scale-invariant $\cP_\zeta$, that is, the dependence of these peaks/valleys's positions on $\alpha$ and $\gamma$ are approximately the same, and we can also observe two notable valleys at $\kappa\sim\cO(1)$ and $\kappa\gtrsim\cO(10)$. 
Obviously, these peaks/valleys also arise from the oscillation of \ac{GW} sources within their dissipation-induced lifetimes. 
Since $\Omega_\mathrm{gw}$ in this ``peak region'' is mainly contributed by $u,v \sim1$, we can conduct an analysis almost identical to that in \cref{sec:3.1} using $\omega_\mathrm{gw}^{\kappa_D\rightarrow\infty}(1,1,\kappa,\hat{\tau}_1,\hat{\tau}_2)$, except replacing the scale-invariant $\cP_\zeta$ by the log-normal one. 
For a relatively wide $\cP_\zeta$, this replacement does not significantly change of the peaks/valleys's positions, but only adjusts their shapes. 

(ii) If dissipative effects happen at the ``infrared region'' (i.e.,  $\kappa_\zeta\gg 1$), the valley at $\kappa\sim\cO(1)$ may not appear (but the valley at $\kappa\sim\cO(10)$ still exists). 
Instead, the infrared behavior $\Omega_\mathrm{gw}^{\kappa_D\rightarrow \infty}\sim  \kappa^3 \ln \kappa^2$ is replaced by a steeper $ \Omega_\mathrm{gw}\sim \kappa^3$ running at $\kappa\lesssim1$. 
The origin of the disappearance of this logarithmic running is the same as discussed in \cref{sec:3.2}, that is, considering the finite lifetimes of \ac{GW} sources, the logarithmic term and the cosine-integral term in the $\cI(u\gg1,v\gg1,\kappa<1\ll \kappa_\zeta,\hat{\tau}\rightarrow\hat{\tau}_l)$ would cancel each other out. 
It is noted that the infrared behavior of $k^3$ is usually regarded as a universal property of the \ac{SGWB}, as required by causality \cite{Cai:2019cdl}.
After taking dissipative effects into account, our results, to a certain extent, restore this universal property. 

\begin{figure*}[t]
    \centering
    \includegraphics[width=\textwidth]{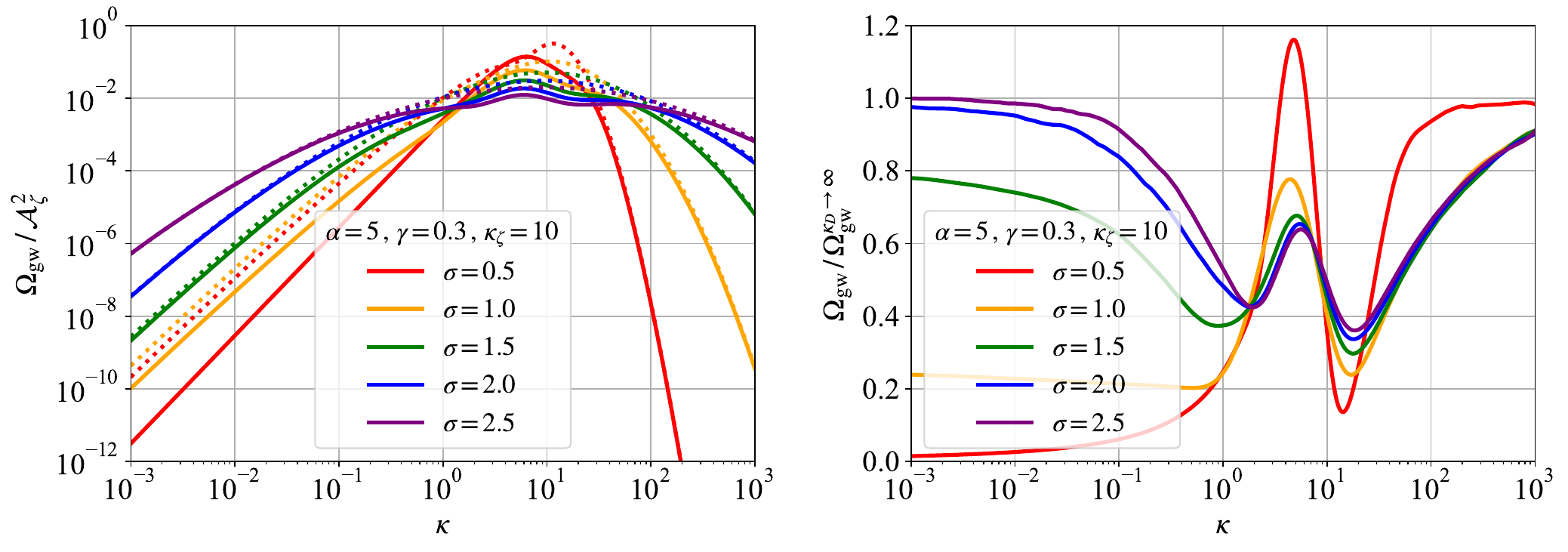}
    \caption{Energy-density spectrum $\Omega_\mathrm{gw}$ normalized by $\cA_\zeta^2$ (left panel) and normalized by $\Omega_\mathrm{gw}^{\kappa_D\rightarrow \infty}$) (right panel) versus the scale $\kappa$ for log-normal primordial curvature power spectra. 
    In both panels, we fix $\alpha=5$, $\gamma=0.3$, $\kappa_\zeta=10$, and take $\sigma=0.5,1.0,1,5,2.0,2.5$. }
    \label{fig:LN_sigma}
\end{figure*}

(iii) If dissipative effects happen at the ``intermediate region'' between the above two, they only suppress the amplitude of $\Omega_\mathrm{gw}$ at $\kappa\lesssim1$, without altering its logarithmic running behavior. 
Different from (i) and (ii), in this region, the \acp{IGW} are not simply dominated by $u,v\sim1$ or $u,v\gg1$, but the contributions of both cannot be ignored. 
In \cref{fig:LN_gamma_kappac} and \cref{fig:LN_alpha_kappac}, row 1 to 4 show the gradual transition process from the ``peak region" to ``infrared region", and the case $\kappa_\zeta \sim 10$ roughly corresponds to the ``intermediate region". 
Besides, we should also note that the specific scope of the ``intermediate region" should be determined by $\cP_\zeta$. 
To demonstrate this, we plot in \cref{fig:LN_sigma} the dissipation effects for different spectral widths $\sigma$, fixing $\kappa_\zeta =10$, $\alpha=5$, and $\gamma=0.3$. 
For a wider $\cP_\zeta$ (e.g., $\sigma=2.5$), the case $\kappa_\zeta =10$ corresponds to the typical ``peak region", while for a narrower $\cP_\zeta$ (e.g., $\sigma=0.5$), it belongs to the ``infrared region". 
These results reflect that dissipative effects in \acp{IGW} have a considerable dependence on the shape of $\cP_\zeta$ itself. 
\\

At the end of this section, we summarize the basic properties of dissipation effects in \acp{IGW} as follows, based on the study of the above specific examples. 
Dissipative effects are most significant around $\kappa \sim 1$, corresponding to the horizon scale at the decoupling of weakly-interacting particles. 
Dissipative effects can leave characteristic imprints on the \ac{IGW} spectrum $\Omega_\mathrm{gw}$, including a notable suppression at $\kappa\sim 1$ and a modification of infrared behavior at $\kappa\lesssim 1$. 
The specific phenomenology largely depends on the shape of $\cP_\zeta$ near $\kappa \sim 1$, and is also affected by the dissipation parameters $\alpha$ and $\gamma$. 
For a relatively flat $\cP_\zeta$, the main effect of dissipation is to cause a suppression with a typical double-valley structure at $\kappa\sim 1$. 
This structure can be regarded as the characteristic that distinguishes dissipative effects from other physical mechanisms. 
Besides, the positions and amplitudes of the valleys are related to $\alpha$ and $\gamma$. 
For a narrow $\cP_\zeta$, since $\Omega_\mathrm{gw}$ is greatly influenced by $\cP_\zeta$ itself, dissipation effects at $\kappa\gtrsim 1$ are usually more complex and model-dependent (In an extreme of monochromatic $\cP_\zeta$, dissipation effects smooth out the resonant peaks/valleys on $\Omega_\mathrm{gw}$). 
However, a relatively universal property of dissipation effects is that, they can remove the logarithmic running of the infrared behavior, replacing it by a pure power-law running. 
This infrared behavior is independent of the shape of $\cP_\zeta$ and the dissipation parameters, but the onset scale at which the logarithmic running disappear is related to them. 
This property can also be regarded as a typical feature of the dissipative effect.
From a physical perspective, the above results stem from the interaction between the dynamics of \ac{GW} sources and dissipation effects, and can be intuitively explained by the ``dissipation-induced lifetime'' of \ac{GW} sources.

\section{Physical implications}\label{sec:4}
Dissipation effects could introduce rich phenomenology on \ac{IGW} spectrum, providing a new window into understanding the early universe. 
This section mainly discusses the implications of dissipative effects in the following two aspects. 
In \cref{sec:4.1}, we will discuss the significant impacts of dissipation effects on \ac{GW} observations. 
We emphasize that dissipation, as an intrinsic nature of cosmic fluid, is a basic factor to be considered in future \ac{GW} analyses. 
In \cref{sec:4.2}, we will focus on the connection between dissipative effects and particle interactions. 
We propose that dissipative effects in \acp{IGW} can provide a new framework for detecting fundamental physics at extremely high energy scales. 

\subsection{For gravitational-wave observations}\label{sec:4.1}

To contact our theoretical results to \ac{GW} observations, we need to know the present-day energy-density fraction spectrum $\Omega_{\mathrm{gw},0}$.
It can be obtained from $\Omega_\mathrm{gw}$ after considering the redshift of radiation components and the change of the relativistic degrees of freedoms in the early Universe, i.e., 
\begin{equation}\label{eq:ogw0}
   \Omega_{\mathrm{gw},0} (f) =
   \Omega_{\mathrm{rad}, 0}\,
   \left[\frac{g_{\ast,\rho}(T_\mathrm{e})}{g_{\ast,\rho}(T_\mathrm{eq})}\right]\, 
   \left[\frac{g_{\ast,s}(T_\mathrm{eq})}{g_{\ast,s}(T_\mathrm{e})}\right]^{4/3}\, 
   \Omega_\mathrm{gw} (k)\big|_{k\simeq2\pi f}\ .
\end{equation}
Here, $f\simeq k/(2\pi)$ is the \ac{GW} frequency, $\Omega_{\mathrm{rad},0}=4.2\times 10^{-5}/h^2$ is the present-day energy-density fraction of radiations with $h=0.674$ the dimensionless Hubble constant \cite{Planck:2018vyg}, $g_{\ast,\rho}$ (or $g_{\ast,\rho}$) is the effective relativistic degrees of freedom associated with the energy density (or entropy), and $T_\mathrm{e}$ (or $T_\mathrm{eq}$) is the cosmic temperature at the \ac{GW} emission (or matter-radiation equality) \cite{Wang:2019kaf}.

As the most basic observable for \acp{IGW}, $\Omega_{\mathrm{gw},0}$ plays a central role in the analysis of observational data. 
Therefore, the precise theoretical calculation of $\Omega_{\mathrm {gw},0}$ is of essential importance to connect \ac{GW} observables with the early Universe. 
This calculation inevitably requires us to take dissipation effects into account in the study of \acp{IGW}. 
This is because that dissipative effects not only universally exist in cosmic fluid, but also become particularly significant near the decoupling of weakly-interacting particles, greatly affecting $\Omega_{\mathrm {gw},0}$ in the relevant frequency bands. 
Our study of dissipative effects provide a more realistic template for $\Omega_{\mathrm {gw},0}$, which has important implications for future \ac{GW} observations, as discussed in the following. 

\begin{figure*}[t]
    \centering
    \includegraphics[width=\textwidth]{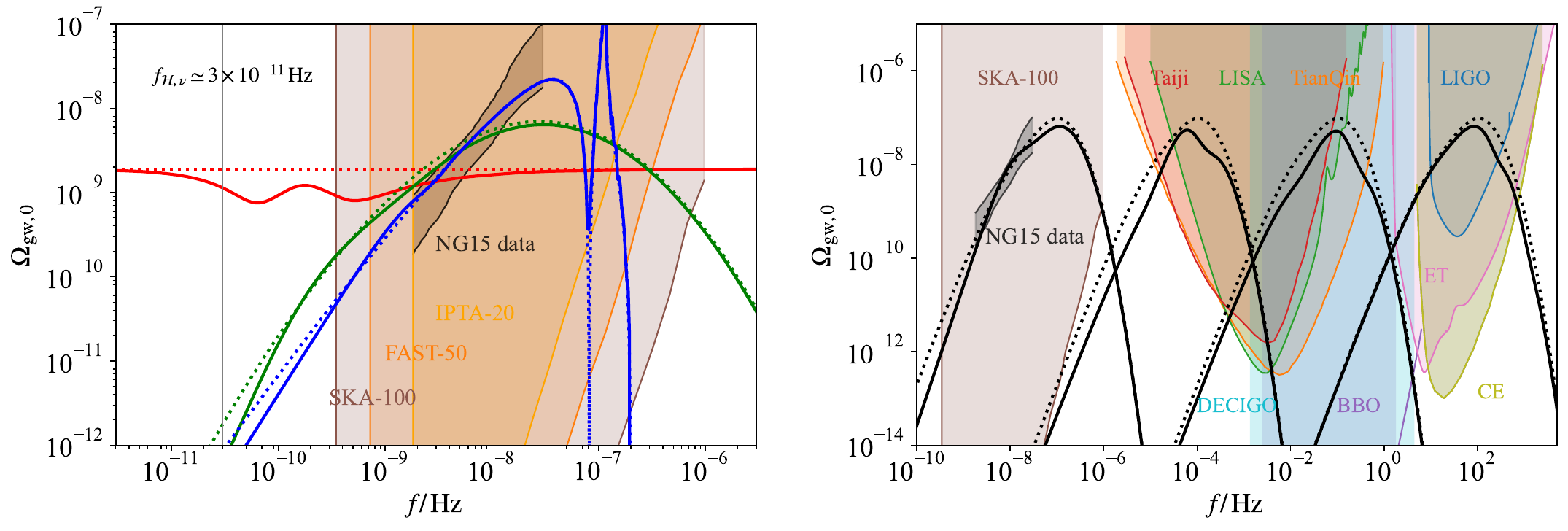}
    \caption{Present-day energy-density fraction spectrum $\Omega_{\mathrm{gw},0}$ with (solid lines) and without (dotted lines) dissipation versus the \ac{GW} frequency. 
    Left panel: 
    The red, blue, and green lines represent the results for scale-invariant $\cP_\zeta$ (with $\cA_\zeta=0.005$), monochromatic $\cP_\zeta$ (with $\cA_\zeta=0.025$ and $f_\zeta=10^{-7}\, \mathrm{Hz}$), and log-normal $\cP_\zeta$ (with $\cA_\zeta=0.05$, $\sigma=2$, and $f_\zeta=3\times10^{-8}\, \mathrm{Hz}$), respectively. 
    Dissipation is caused by neutrinos, which is characterized by $\alpha\simeq5$, $\gamma\simeq 0.3$, and $f_{\cH,\nu}\simeq 3\times 10^{-11}\, \mathrm{Hz}$. 
    Right panel: 
    The black lines represent the results for log-normal $\cP_\zeta$ with $\cA_\zeta=0.1$, $\sigma=1$, and $f_{\zeta}=10^{-7},10^{-4},10^{-1},10^{2}\, \mathrm{Hz}$ (from left to right). 
    Dissipation is caused by possible new weakly-interacting particles $X$, where we set $\alpha\simeq5$, $\gamma\simeq 0.3$, and $f_{\zeta}/f_{\cH,X}=100,10,5,1$ (from left to right).}
    \label{fig:Dissipation_obs}
\end{figure*}

Firstly, let us focus on the dissipative effects within the \ac{SM}, which are dominated by weak-interaction of neutrinos, as formulated in \cref{eq:specific cross section}.
In this case, the dissipation parameters are given by $\alpha\simeq5$ and $\gamma\simeq 0.3$, and the frequency corresponding to the horizon scale at neutrino decoupling is estimated as $f_{\cH,\nu}\simeq k_{\cH,\nu}/(2\pi) \simeq 3\times 10^{-11}\, \mathrm{Hz}$. 
In the left panel of \cref{fig:Dissipation_obs}, we plot the results of $\Omega_{\mathrm{gw},0}$ with (solid lines) and without (dotted lines) dissipative effects, and compare them with the sensitivity curvies of IPTA (\ac{NANOGrav} \cite{Jenet:2009hk}, European Pulsar Timing Array (EPTA) \cite{Kramer:2013kea}, and Parkes Pulsar Timing Array (PPTA) \cite{Manchester:2012za} combined), Five-Hundred-Meter Aperture Spherical Radio Telescope (FAST) \cite{Nan:2011um}, and \ac{SKA} \cite{2009IEEEP..97.1482D}) for a 20-, 50-, and 100-year observation \cite{Kuroda:2015owv}, respectively. 
As a result, dissipative effects would become significant around nano-hertz band, which will be an important observational objective for \ac{PTA} experiments.
For example, for a scale-invariant $\cP_\zeta=0.005$ (red solid line), the corresponding $\Omega_{\mathrm{gw},0}$ will be suppressed dissipation effects by several tens of percent within the \ac{PTA} frequency band (maximally by $\simeq 60 \%$ at $f\simeq 0.5\, \mathrm{nHz}$). 
In particular, the characteristic valley structure at $f\simeq 0.5\, \mathrm{nHz}$ caused by neutrino decoupling can also be detected in future \ac{PTA} experiments. 
Once detected, it will not only verify dissipative effects in \acp{IGW}, but also provide the unique feature for \acp{IGW} that distinguish them from other potential nanohertz \ac{GW} sources (e.g., supermassive black hole binaries), helping us differentiate the components of \ac{SGWB}. 
Our results indicate that, even within the \ac{SM}, dissipative effects are of great significance to \ac{GW} observations. 

Since \acp{IGW} are the candidate for the \ac{GW} signals reported by \acp{PTA} \cite{NANOGrav:2023hvm}, it is essential to consider the impact of dissipation effects on \ac{PTA} data analysis. 
As an illustration, in the left panel of \cref{fig:Dissipation_obs}, we plot the blue and green lines as possible interpretations of the \ac{NG15} data \cite{NANOGrav:2023gor} (gray region), where the blue (or green) lines correspond to the monochromatic (or log-normal) $\cP_\zeta$ with $\cA_\zeta=0.025$ and $f_\zeta=10^{-7}\, \mathrm{Hz}$ (or, $\cA_\zeta=0.05$, $\sigma=2$, and $f_\zeta=3\times10^{-8}\, \mathrm{Hz}$). 
For the narrow $\cP_\zeta$,  dissipation effects caused bu neutrino decoupling happen in the far-infrared region.
The frequency band at which they significantly modify the infrared behavior of $\Omega_{\mathrm{gw},0}$ is slightly below the nanohertz band, making it difficult to be observed in current \ac{PTA} experiments. 
However, for the wide $\cP_\zeta$, we find that these dissipative effects could visibly change $\Omega_{\mathrm{gw},0}$ at $f\gtrsim \mathrm{nHz}$, thereby affecting the analysis of \ac{GW} data. 
We leave the quantitative discussion to the future work. 
As the observation duration increases, it is expected that the \acp{PTA} experiment will detect \acp{GW} at lower frequencies, thus enabling a more comprehensive study of these dissipative effects caused by neutrinos.

Furthermore, we can consider the dissipative effects caused by underlying new physics. 
If there exist new paritcles $X$ with weaker interaction than neutrinos, they would decouple earlier and the corresponding dissipative effects happen at higher frequencies. 
In the right panel of \cref{fig:Dissipation_obs}, we plot the results of $\Omega_{\mathrm{gw},0}$ for a log-normal $\cP_\zeta$ with (solid lines) and without (dotted lines) dissipative effects, where we take $f_{\zeta}=10^{-7},10^{-4},10^{-1},10^{2}\, \mathrm{Hz}$ and $f_{\zeta}/f_{\cH,X}=100,10,5,1$ from left to right, and fix $\cA_\zeta=0.1$, $\sigma=1$, $\alpha=5$, and $\gamma=0.3$. 
In these case, dissipative effects in \acp{IGW}, including the characteristic suppression and the transition of infrared behavior, are expected to be detected by a wide range of \ac{GW} experiments, such as \ac{PTA} experiments (e.g., \ac{NANOGrav} \cite{Jenet:2009hk} and \ac{SKA} \cite{2009IEEEP..97.1482D}), space-based \ac{GW} interferometers (e.g., \ac{LISA} \cite{LISA:2017pwj,2019BAAS...51g..77T,LISACosmologyWorkingGroup:2022jok}, Taiji \cite{Hu:2017mde}, TianQin \cite{TianQin:2015yph}, \ac{DECIGO} \cite{Seto:2001qf,Kawamura:2020pcg}, \ac{BBO} \cite{Crowder:2005nr,Smith:2016jqs}), and
ground-based \ac{GW} interferometers (e.g., LIGO \cite{Harry_2010}, \ac{ET} \cite{Punturo:2010zz}, and \ac{CE} \cite{Reitze:2019iox}).
Their implications for new physics searches will be discussed in \cref{sec:4.2}. 

The infrared behavior of $\Omega_{\mathrm {gw},0}$ is the key information of \ac{SGWB}. 
As for \acp{IGW}, their universal logarithmic infrared running is regarded as a unique feature to distinguish them from other sources \cite{Yuan:2019wwo,Cai:2019cdl,Cai:2018dig,Adshead:2021hnm,Han:2025wlo,Luo:2025lgr,Li:2025met}. 
This property actually comes from the sufficiently long lifetimes of the sources of \acp{IGW}. 
However, the situation becomes more complex in the presence of dissipation. 
Roughly speaking, this logarithmic running remains unchanged if dissipative effects happen in the ``peak region'', but disappers if they happen in the typical ``infrared region'', As stated in \cref{sec:3}. 
Moreover, a complete analysis of the infrared behavior will be related to both the shape $\cP_\zeta$ and the dissipation parameters. 
Our work emphasizes that, to describe the infrared behavior of \acp{IGW} more realistically, it is necessary to consider dissipative effects. 

In addition to reshaping the profile of $\Omega_{\mathrm{gw},0}$, dissipative effects inherently lead to a reduction in the total energy density of \acp{IGW}, i.e., $\int \Omega_{\mathrm{gw},0}(f)\, \ud\ln{f}$. 
For instance, in the case of a log-normal $\cP_\zeta$ with $\sigma=1$, the total \ac{GW} energy fraction in the dissipation-free case is $\simeq 2.03 \times 10^{-5}\, \cA^2_\zeta$. 
When dissipative effects are included (e.g., with $\alpha=5$, $\gamma=0.3$, and $f_\zeta/f_{\cH,j}=1$), this value is reduced to $\simeq 1.31 \times 10^{-5}\, \cA^2_\zeta$. 
Such reduction has noteworthy cosmological implications, as the big-bang nucleosynthesis and the \ac{CMB} constrain the total energy density of  extra relativistic species like \acp{GW} \cite{Smith:2006nka,Clarke:2020bil} and therefore set bounds on $\cA_\zeta$. 
As a result, dissipative effevts can release these bounds compared to the dissipation-free case.

\subsection{For new physics searches}\label{sec:4.2}

An important application of dissipative effects in \acp{IGW} is their use as a probe for new physics, owing to their close connection with particle interactions in the early Universe.  
Despite the \ac{SM}'s tremendous success in describing known particle interactions, unresolved issues such as neutrino masses and the matter-antimatter asymmetry strongly indicate the existence of the \ac{BSM} physics.  
A common feature of many \ac{BSM} scenarios is the introduction of extended gauge symmetries broken spontaneously at high energy scales, typically accompanied by additional heavy gauge bosons, whose masses may span many orders of magnitude.
For instance, the \ac{GUT} with gauge groups larger than $SU(5)$ generally predict additional gauge bosons, whose masses are allowed to range from the electroweak scale up to the \ac{GUT} scale \cite{Leike:1998wr}.  
The detection of these heavy gauge bosons has essential implications for validating new physics scenarios.
Yet, current and near-future collider experiments face significant challenges accessing energy scales well beyond the TeV range.

The early Universe, as a natural high-energy laboratory, offers a unique observational window to explore new physics.  
If new weakly-interacting particles $X$ exist, their interactions with the cosmic plasma are expected to induce sizable dissipation upon decoupling in the early Universe. 
This dissipation is characterized by a damping scale $k_{D,X}$, which encodes key information about the underlying particle interactions.
Correspondingly, these dissipative processes imprint distinctive signatures in the IGW spectrum at scales $k\sim k_{D,X}$, presenting detectable targets for future \ac{GW} observations. 
Hence, dissipative effects in \acp{IGW} provide an indirect probe of new physics, especially at energy scales beyond the reach of traditional particle physics experiments.
Building upon the initial proposal in Ref.~\cite{Yu:2024xmz}, this work further develops this approach by enhancing the phenomenological understanding of dissipative effects in \acp{IGW}. 
For a broad $\cP_\zeta$, a characteristic suppression on $\Omega_\mathrm{gw}$, especially a multi-peak structure, if detected, serves as the fingerprint of the decoupling of some underlying weakly-interacting particles $X$, providing a smoking gun for new physics. 
The frequency location of this feature reveals the time of $X$-particle decoupling, which encodes the crucial information about their interactions. 
Moreover, the detailed shape of these spectral features enables us to extract the dissipation parameters $\alpha$ and $\gamma$, providing additional implications on the particle abundances and interactions. 
For a narrow $\cP_\zeta$, identifying dissipative effects is usually more complex because $\Omega_\mathrm{gw}$ is largely shaped by $\cP_\zeta$ itself. 
Nevertheless, the transition from $\sim k^n\ln{k^2}$ to $\sim k^n$ ($n=2,3$) in the infrared region could be a relatively general property in the presence of dissipation. 
Once detected, it also serves as typical signals of dissipative effects, suggesting the existence of new weakly-interacting particles. 
In the following, we mainly discuss on the former case. 

To quantitatively present the detection potential to new physics at extremely high energy scales, we consider the interaction of weakly-interacting particles $X$ parameterized by \cref{eq:specific cross section}. 
In this example, the mass of the mediated gauge bosons, $M_{W'}$, corresponds to the energy scale of the related new physics. 
When these $X$-particles decouple in the early Universe, the damping scale is roughly estimated as $k_{D,X}/\mathrm{Mpc}^{-1} \sim  10^2\,\left(M_{W'}/\mathrm{GeV}\right)^{4/3}$, and the associated dissipative effects on $\Omega_\mathrm{gw,0}(f)$ occurs at frequencies approximately 
\begin{equation}\label{eq:fM'}
    {f}/{\mathrm{Hz}}\sim \cO(1-10)\times 10^{-13}\,\left({M_{W'}}/{\mathrm{GeV}}\right)^{4/3}\ .
\end{equation}
Based on \cref{eq:fM'}, \ac{PTA} experiments are projected to detect dissipative effects corresponding to the parameter space $M_{W'} \sim \mathcal{O}(10^{3} - 10^{4})\, \mathrm{GeV}$. 
This energy scale range is particularly noteworthy as it overlaps with the anticipated sensitivity of next-generation colliders, offering valuable cross-check with experiments such as Large Hadron Collider (LHC), Future Circular Collider (FCC) \cite{FCC:2018evy}, International Linear Collider (ILC) \cite{ILCInternationalDevelopmentTeam:2022izu}, and Circular Electron Positron Collider (CEPC) \cite{CEPCStudyGroup:2018rmc,CEPCStudyGroup:2018ghi,Antusch:2025lpm}.
Space-based gravitational wave interferometers probe higher energy scales, corresponding to $M_{W'} \sim \mathcal{O}(10^{7} - 10^{9})\, \mathrm{GeV}$, while ground-based \ac{GW} interferometers extend the sensitivity up to $M_{W'} \sim \mathcal{O}(10^{10} - 10^{12})\, \mathrm{GeV}$. 
These scales far exceed the reach of terrestrial collider experiments, yet offer valuable opportunities to explore new physics through cosmic observations. 
Complementary observations come from high-energy cosmic ray experiments, such as Pierre Auger Observatory (PAO) \cite{PierreAuger:2015eyc}, Telescope Array (TA) \cite{TelescopeArray:2008toq}, and Large High Altitude Air Shower Observatory (LHAASO) \cite{LHAASO:2019qtb}.
We expect that such multi-messenger search could scan a wide range of energy scales, enriching our capability to explore new physics. 

We note that since $\cP_\zeta$ is unknown a priori, there may be some uncertainties in the realistic analysis of dissipative effects in \acp{IGW}. 
However, we would like to emphasize the robustness of this approach by discussing the following aspects about $\cP_\zeta$. 
\begin{itemize}
 \item Position of $k_\zeta$: 
To detect dissipative effects, we at least require large enough $\cP_\zeta$ at relevant scales to generate observable \acp{IGW}. 
For such $\cP_\zeta$, the relative position between its typical scale $k_\zeta$ and damping scale $k_{D,X}$ significantly influences the manifestation of dissipative effects in \acp{IGW}. 
Obviously, the suppression in $\Omega_\mathrm{gw}$ is more pronounced when $k_\zeta$ is closer to $k_{D,X}$. 
However, we observe that even if $k_\zeta$ and $k_{D,X}$ differ significantly (e.g., $k_\zeta/k_{D,X} \sim 10^2$), dissipative effects in \acp{IGW} remain notable enough, reducing $\Omega_\mathrm{gw}$ by several tens of percent, as shown in \cref{sec:3.3}.
This indicates that the detection of dissipative effects is feasible across a wide parameter space, without requiring exact scale matching. 
 \item Shape of $\cP_\zeta$: 
Though lacking the specific shape of $\cP_\zeta$ makes it difficult to precisely measure dissipative effects, we emphasize that for our purpose of probing new physics, it is usually not a serious problem. 
This is because in our approach, the most robust and accessible observable is the feature's frequency, instead of its exact shape. 
As long as the dissipation-induced feature is identifiable, its frequency can be well measured, from which we can extract the particle interaction information even without the knowledge about $\cP_\zeta$ (as shown in \cref{eq:fM'}). 
Moreover, the shape of dissipation-induced feature is more likely to be well measured for a relatively broad $\cP_\zeta$, since it is less affected by $\cP_\zeta$ itself. 
In this case, we can extract dissipation parameters $\alpha$ and $\gamma$ and get further implications for the related new physics, e.g., the abundance of $X$-particles. 
  \item Possible degeneracy: 
Features in $\Omega_\mathrm{gw}$ consistently attract significant attention as they strongly indicate the presence of nontrivial physical processes.  
As for \acp{IGW}, both dissipative effects and $\cP_\zeta$ itself can generate features in $\Omega_\mathrm{gw}$ (for the latter, see Ref.~\cite{Braglia:2020taf} for a review).  
A natural question arises whether features from these two origins can be degenerate.  
Though the degeneracy cannot be theoretically ruled out, we emphasize that the unique characteristics of dissipation-induced features, most notably their multi-peak structure, distinguish them from features caused by other known physical scenarios, offering a promising means to identify particle decoupling in the early Universe.  
Furthermore, given that degeneracies often appear in \ac{GW} cosmology, this motivates future research to incorporate complementary observables, such as \ac{GW} anisotropies, which may help break potential degeneracies.
\end{itemize}

Although the generation of $\cP_\zeta$ and the presence of new particles are usually independent, it is interesting that some cosmological models predict both simultaneously. 
For example, Ref.~\cite{Kawasaki:2019hvt} considers a $U(1)$ gauge field kinetically coupled to an inflaton, which could copiously produce new heavy gauge bosons, and also enhance $\cP_\zeta$ to $\sim 10^{-3}$. 
If the damping scale is determined by the gauge boson's mass through \cref{eq:fM'} and has overlap with the enhanced scales of $\cP_\zeta$, the corresponding dissipative effects in \acp{IGW} are potentially detected by \ac{GW} experiments. 
A detection of such effects could probe these new particles, while the non-detection can restrict the parameter space, providing an important test of this model. 
In principle, given such a new physics model, we can calculate the corresponding dissipative effects in \acp{IGW} following our flamework, and compare this template to the \ac{GW} observations. 
Our work may lay the foundation for future model-building efforts combining enhanced primordial perturbations and new particle content, an endeavor valuable both theoretically and observationally, which strengthens the interface between particle physics and cosmology.

In summary, dissipative effects in \acp{IGW} serve as characteristic features of the decoupling of weakly-interacting particles and encode rich information on their interactions. 
These effects provide a new framework to probe new physics at extremely high energy scales, advancing our understanding of the early Universe and fundamental physics.

\section{Extended study}\label{sec:5}

We have investigated dissipative effects in \acp{IGW} in the most standard case, that is, we assume primordial curvature perturbations $\zeta$ to be Gaussian, and only consider the energy-density spectrum of \acp{IGW} produced in the \ac{RD} era.
As an extended study, we here discuss some basic contents about the following three topics, while leave their detailed researches to future work. 
In \cref{sec:5.1}, we study the dissipative effects in \acp{IGW} in the case that $\zeta$ deviate from Gaussian statistics, which is known as primordial non-Gaussianity. 
In \cref{sec:5.2}, we consider the anisotropies in the \ac{IGW} background caused by primordial non-Gaussianity, and study the role of dissipation here. 
In \cref{sec:5.3}, we turn to \acp{IGW} related to the \ac{eMD} era, and focus how dissipation affects the so-called ``poltergeist mechanism''.

\subsection{Dissipative effects in the presence of primordial non-Gaussianity}\label{sec:5.1}

While the standard single-field slow-roll inflation predicts nearly Gaussian $\zeta$, a broad range of inflation models (e.g., non-standard single-field inflation \cite{Namjoo:2012aa,Martin:2012pe,Chen:2013aj,Huang:2013oya,Mooij:2015yka,Bravo:2017wyw,Finelli:2017fml,Cai:2018dkf,Passaglia:2018ixg} or multi-field inflation \cite{Lyth:2002my,Bartolo:2003jx,Zaldarriaga:2003my,Lyth:2005qk,Frazer:2011br,Linde:2012bt,McAllister:2012am,Torrado:2017qtr,Bjorkmo:2017nzd}) can generate sizable primordial non-Gaussianity, which reflects the interactions of quantum fields during inflation. 
In this work, we consider the local-type non-Gaussianity, and expand $\zeta$ up to the $F_{NL}$-order as follows
\begin{equation}\label{eq:FNL}
    \zeta_\bk 
    = \zeta_{g,\bk} 
    +  F_\mathrm{NL} 
    \int \frac{\ud^3 \bq}{(2\pi)^{3/2}} \ 
    \zeta_{g,\bk-\bq} \,\zeta_{g,\bq}\ ,
\end{equation}
where $\zeta_g$ is the Gaussian component of $\zeta$. 
Given the non-Gaussian $\zeta$ in \cref{eq:FNL}, the \ac{IGW} spectrum ${\Omega}_\mathrm{gw} \sim \langle \zeta^4\rangle $ can be expanded in terms of $\langle \zeta_g^2\rangle$ and finally recast as ${\Omega}_\mathrm{gw} = {\Omega}_\mathrm{gw} ^{(0)}+{\Omega}_\mathrm{gw} ^{(1)}+{\Omega}_\mathrm{gw} ^{(2)}$, where ${\Omega}_\mathrm{gw} ^{(m)}$ ($m=0,1,2$) stand for the  $\cO(\cA_\zeta^{m+2} F_\mathrm{NL}^{2m})$ components of ${\Omega}_\mathrm{gw}$. 
The explicit expressions of ${\Omega}_\mathrm{gw} ^{(m)}$ are given by \cite{Adshead:2021hnm,Ragavendra:2021qdu,Li:2023qua}
\begin{eqnarray}\label{eq:Omega-n}
    {\Omega}_\mathrm{gw}^{(0)} (\kappa,\hat{\tau})
    &=& \frac{1}{12}
        \int_{-1}^1 \ud s \int_0^\infty \ud t \ 
        u v\, 
        \overbar{J^2 (u,v,\kappa,\hat{\tau})}\ 
        \frac{\cP_{\zeta_g}(v\kappa)}{v^3}
        \frac{\cP_{\zeta_g}(u\kappa)}{u^3} \ ,
    \label{eq:Omega-0}
    \\
        \nonumber \\ 
    {\Omega}_\mathrm{gw}^{(1)} (\kappa,\hat{\tau})
    &=& \frac{1}{12}\, F_\mathrm{NL}^{2}
        \int_{-1}^1 \ud s_1 \int_0^\infty \ud t_1 
        \int_{-1}^1 \ud s_2 \int_0^\infty \ud t_2 
        \nonumber\\
        & &\times\Bigg\{
        u_1 v_1^4 u_2 v_2\, 
        \overbar{J^2 (u_1,v_1,\kappa,\hat{\tau})}\ 
        \frac{\cP_{\zeta_g}(u_1\kappa)}{u_1^3}
        \frac{\cP_{\zeta_g}(v_2 v_1\kappa)}{(v_2 v_1)^3}
        \frac{\cP_{\zeta_g}(u_2 v_1\kappa)}{(u_2 v_1)^3} \nonumber\\
        & &\quad
        +u_1 v_1 u_2 v_2\, 
        \overbar{J (u_1,v_1,\kappa,\hat{\tau})\, 
        J (u_2,v_2,\kappa,\hat{\tau})}\times
        \int_0^{2\pi} \frac{\ud \varphi_{12}}{\pi} \,\cos{2\varphi_{12}} 
        \nonumber\\
        & &\quad
        \times\Bigg[
        \frac{\cP_{\zeta_g}(v_2\kappa)}{v_2^3}
        \frac{\cP_{\zeta_g}(u_2\kappa)}{u_2^3}
        \frac{\cP_{\zeta_g}(w_{12}\kappa)}{w_{12}^3}
        +
        \frac{\cP_{\zeta_g}(v_1\kappa)}{v_1^3}
        \frac{\cP_{\zeta_g}(v_2\kappa)}{v_2^3}
        \frac{\cP_{\zeta_g}(w_{012}\kappa)}{w_{012}^3}
        \Bigg]\Bigg\}
        \ ,
    \label{eq:Omega-1}
    \\
            \nonumber \\ 
    {\Omega}_\mathrm{gw}^{(2)} (\kappa,\hat{\tau})
    &=& \frac{1}{48}\, F_\mathrm{NL}^{4}
        \int_{-1}^1 \ud s_1 \int_0^\infty \ud t_1 
        \int_{-1}^1 \ud s_2 \int_0^\infty \ud t_2
        \int_{-1}^1 \ud s_3 \int_0^\infty \ud t_3 
        \nonumber\\
        & &\times\Bigg\{
        u_1^4 v_1^4 u_2 v_2 u_3 v_3\, 
        \overbar{J^2 (u_1,v_1,\kappa,\hat{\tau})}\ 
        \frac{\cP_{\zeta_g}(v_2 v_1\kappa)}{(v_2 v_1)^3}
        \frac{\cP_{\zeta_g}(u_2 v_1\kappa)}{(u_2 v_1)^3}
        \frac{\cP_{\zeta_g}(v_3 u_1\kappa)}{(v_3 u_1)^3}
        \frac{\cP_{\zeta_g}(u_3 u_1\kappa)}{(u_3 u_1)^3} \nonumber\\
        & &\quad
        +u_1 v_1 u_2 v_2 u_3 v_3\, 
        \overbar{J (u_1,v_1,\kappa,\hat{\tau})\, 
        J (u_2,v_2,\kappa,\hat{\tau})}\times
        \frac{1}{2}
        \int_0^{2\pi} \frac{\ud \varphi_{12}}{\pi}
        \int_0^{2\pi} \frac{\ud \varphi_{23}}{\pi}
        \,\cos{2\varphi_{12}}
        \nonumber\\
        & &\quad
        \times
        \frac{\cP_{\zeta_g}(v_3\kappa)}{v_3^3}
        \frac{\cP_{\zeta_g}(w_{13}\kappa)}{w_{13}^3}
        \frac{\cP_{\zeta_g}(w_{23}\kappa)}{w_{23}^3}
        \Bigg[
        2\, \frac{\cP_{\zeta_g}(u_3\kappa)}{u_3^3}
        +
        \frac{\cP_{\zeta_g}(w_{0123}\kappa)}{w_{0123}^3}
        \Bigg]\Bigg\}\ ,
    \label{eq:Omega-2}
\end{eqnarray}
with ${J(u,v,\kappa,\hat{\tau})}
    =\left[{v^2-(1+v^2-u^2)^2/4}\right]\times
    {\cI(u,v,\kappa,\hat{\tau})}$
\footnote{
The notations in \cref{eq:Omega-n} are defined as follows
\begin{equation}
    \begin{aligned}
    s_i &= u_i - v_i\ , \\
    t_i &= u_i + v_i -1\ , \\
    y_{i} &= \frac{1}{2}[1-s_i(t_i+1)]\ , \\
    y_{ij} &=  \frac{\cos{\varphi_{ij}}}{4}\sqrt{t_i (t_i + 2) (1 - s_i^2) t_j (t_j + 2) (1 - s_j^2)} 
            + \frac{1}{4}[1 - s_i (t_i + 1)][1 - s_j (t_j + 1)]\ , \\
    w_{ij} &= \sqrt{v_i^2 + v_j^2 - 2 y_{ij}}\ , \\
    w_{012} &= \sqrt{1+v_1^2 + v_2^2 -2 y_{1} - 2y_{2} +2 y_{12}}\ , \\
    w_{0123} &= \sqrt{1+v_1^2 + v_2^2 + v_3^2 - 2y_1 - 2y_2 + 2y_3 + 2 y_{12} - 2 y_{23} - 2 y_{31}}\ .
    \end{aligned}
\end{equation}
}.
Note that \cref{eq:Omega-n} just has the same form as the results in existing literature except that the ${\cI(u,v,\kappa,\hat{\tau})}$ here includes dissipative effects. 

\begin{figure*}[htbp]
    \centering
    \includegraphics[width=1\textwidth]{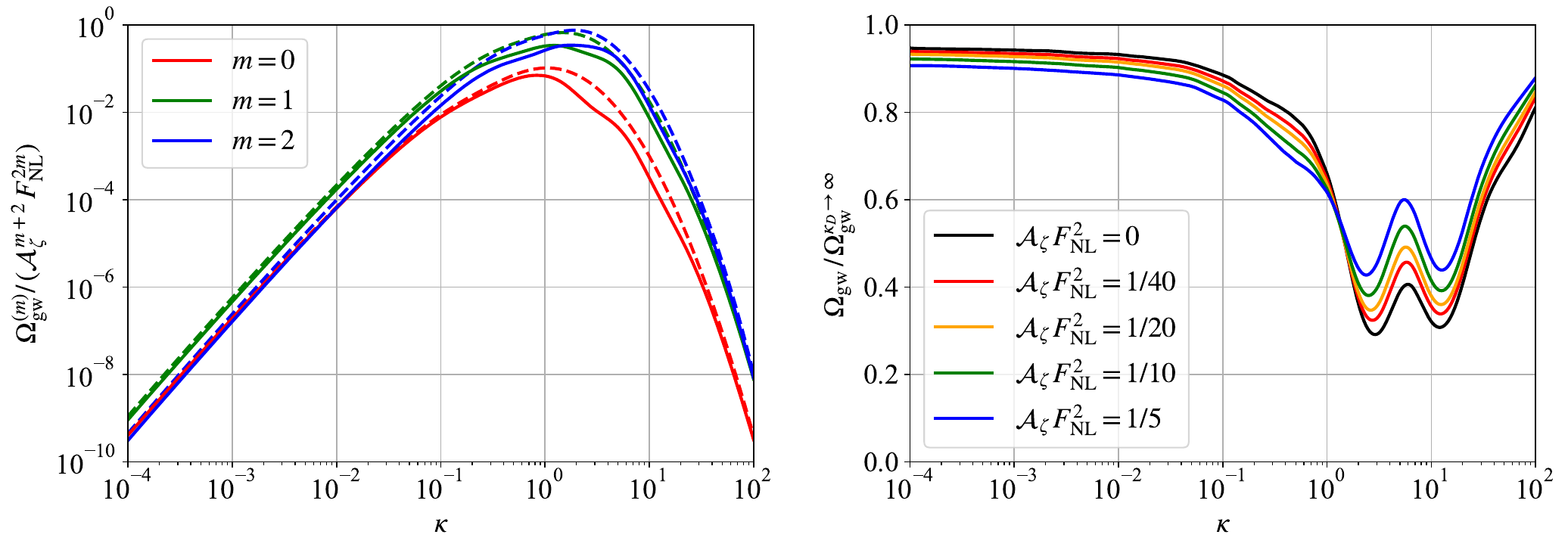}
    \caption{Left panel: Non-Gaussian components of energy-density fraction spectrum, ${\Omega}_\mathrm{gw} ^{(m)}$ (normalized by $\cA_\zeta^{m+2} F_\mathrm{NL}^{2m}$, $m=0,1,2$), versus the scale $\kappa$ with (solid lines) and without (dashed lines) dissipation. 
    Right panel: Total energy-density fraction spectrum ${\Omega}_\mathrm{gw}$ (normalized by the dissipation-free result ${\Omega}^{\kappa_D\rightarrow\infty}_\mathrm{gw}$) versus the scale $\kappa$ for various values of $\cA_\zeta F_\mathrm{NL}^2$. 
    In both panels, we fix the dissipation parameters to $\alpha=5$, $\gamma=0.3$, and assume a log-normal primordial power spectrum with $\sigma=1$ and $\kappa_\zeta=1$ for $\zeta_g$.} 
    \label{fig:PNG}
\end{figure*}

In \cref{fig:PNG}, the left panel plots the numerical results of ${\Omega}_\mathrm{gw} ^{(m)}$ (nomarlized by $\cA_\zeta^{m+2} F^{2m}_\mathrm{NL}$) with or without dissipative effects, and the right panel plots the ratio ${\Omega}_\mathrm{gw}\,/\,{\Omega}^{\kappa_D\rightarrow\infty}_\mathrm{gw}$ for different non-Gaussianity parameters $\cA_\zeta F^2_\mathrm{NL}$. 
Here, we take the dissipation parameters as $\alpha=5$ and $\gamma=0.3$, and assume $\cP_{\zeta_g}(\kappa)$ a log-normal spectrum with $\kappa_\zeta=1$ and $\sigma=1$. 
In the presence of primordial non-Gaussianity, we find that dissipative effects still result in a peak/valley structure on $\Omega_\mathrm{gw}$, whose location remains nearly unchanged comparing to the Gaussian case. 
The effect of non-Gaussianity is to change the contrast ratio of the dissipation-induced peak/valley structure, making it less obvious as the parameter $\cA_\zeta F^2_\mathrm{NL}$ increases. 
It can be intuitively understood as follows: 
Since the non-Gaussianity introduces non-linear couplings between different scales, the damping of the mode $k\sim k_{D,j}$ also affect other modes. 
As a result, dissipative effects on ${\Omega}_\mathrm{gw}$ located at $\kappa\sim \cO(1-10)$ would ``spread out" to other scales, thus reducing the contrast ratio of the features. 
This effect is quite different with those of dissipation parameters themselves, serving as distinctive signals of primordial non-Gaussianity. 

\subsection{Anisotropies in the induced gravitational-wave background}\label{sec:5.2}

Besides the isotropic component of \ac{SGWB} described by $\Omega_\mathrm{gw}$, the anisotropies of \ac{SGWB} are another important \ac{GW} observable and could further bring us crucial information about the early Universe. 
The anisotropies are characterized by the reduced angular power spectrum $\widetilde{C}_\ell$, defined as the correlation of large-scale \ac{GW} density contrasts $\delta_{\mathrm{gw},0,\bk}$ along two lines-of-sight, i.e., 
\begin{equation}\label{eq:Cl-def}
    \delta_{\ell_1 \ell_2} \delta_{m_1 m_2} \widetilde{C}_\ell (k)
    = \langle\delta_{\mathrm{gw},0,\ell_1 m_1}(k)\,\delta_{\mathrm{gw},0,\ell_2 m_2}^\ast(k)\rangle\ .
\end{equation}
Here, $\bk$ is the Fourier mode of \acp{GW}, $\delta_{\mathrm{gw},0,\ell m}$ is the spherical harmonic coefficient of $\delta_{\mathrm{gw},0,\bk}$, namely, $\delta_{\mathrm{gw},0,\bk} = \Sigma_{\ell m}\  \delta_{\mathrm{gw},0,\ell m}(k)\, Y_{\ell m}({\bk/k})$. 
As for \acp{IGW}, the large-scale $\delta_{\mathrm{gw},0,\bk}$ can be naturally generated by primordial non-Gaussianity \cite{Bartolo:2019zvb,LISACosmologyWorkingGroup:2022kbp,LISACosmologyWorkingGroup:2022jok,Schulze:2023ich,Li:2023qua,Li:2023xtl,Wang:2023ost,Yu:2023jrs,Ruiz:2024weh,Rey:2024giu,Luo:2025lgr,Yu:2025jgx,Li:2025met}, since it can redistribute the \ac{IGW} density through the coupling between large- and small-scale curvature perturbations. 
At a coarse-grained location $\bx$, the large-scale $\delta_{\mathrm{gw},0,\bk}$ of \acp{IGW} is given by \cite{Li:2023qua}
\begin{equation}\label{eq:delta}
    \delta_{\mathrm{gw},0,\bk}(\bx) \simeq
    \left\{F_\mathrm{NL}\,
    \frac{8\,{\Omega}_\mathrm{gw}^{(0)}(k)
    +4\,{\Omega}_\mathrm{gw}^{(1)}(k)}{{\Omega}_\mathrm{gw}^{(0)}(k)+{\Omega}_\mathrm{gw}^{(1)}(k)+{\Omega}_\mathrm{gw}^{(2)}(k)} 
    + \frac{3}{5}\,\left[4-n_\mathrm{gw} (k)\right]
    \right\}\times
    \int \frac{\ud^{3}\bq}{(2\pi)^{3/2}}\ 
    e^{i\bq\cdot\bx} \,
    \zeta_{gL,\bq}\ ,
\end{equation}
where $n_\mathrm{gw} (k)  =  {\partial \ln {\Omega}_\mathrm{gw} (k)}\,/\,{\partial\ln k}$ is the spectral index of ${\Omega}_\mathrm{gw}$, and $\zeta_{gL}$ is the large-scale curvature perturbations. 
In the right-hand side of \cref{eq:delta}, the first term is the density contrast of \acp{GW} at their emission arising from primordial non-Gaussianity, while the second term is known as the Sachs-Wolfe effect, corresponding to the redshift/blueshift of \acp{GW} during their propagation due to the existence of large-scale gravitational potential \cite{Sachs:1967er}. 
We here do not include other propagation effects, which are expected to be negligible \cite{Bartolo:2019zvb}. 
Combining \cref{eq:Cl-def} and \cref{eq:delta}, the $\widetilde{C}_\ell$ of the \ac{IGW} background can be obtained as \cite{Li:2023qua}
\begin{equation}\label{eq:Cl}
    \ell (\ell+1)\,\widetilde{C}_\ell (\kappa) 
    = {2\pi \cA_{\mathrm{L}}} 
    \left\{
        F_\mathrm{NL}\,
        \frac{8\,{\Omega}_\mathrm{gw}^{(0)}(\kappa)
        +4\,{\Omega}_\mathrm{gw}^{(1)}(\kappa)}{{\Omega}_\mathrm{gw}^{(0)}(\kappa)+{\Omega}_\mathrm{gw}^{(1)}(\kappa)+{\Omega}_\mathrm{gw}^{(2)}(\kappa)}
        + \frac{3}{5} \left[4 - n_\mathrm{gw} (\kappa)\right] 
    \right\}^2\ ,
\end{equation}
where $\cA_{\mathrm{L}}\simeq 2.1\times 10^{-9}$ is the spectral amplitude of $\zeta_{gL}$ given by \ac{CMB} observations \cite{Planck:2018vyg}. 
In \cref{eq:Cl}, dissipative effects could affect the anisotropies of \acp{IGW} by modifying $\Omega_\mathrm{gw}^{(m)}(\kappa)$ ($m=0,1,2$) as well as $n_\mathrm{gw}(\kappa)$. 

\begin{figure*}[htbp]
    \centering
    \includegraphics[width=1\textwidth]{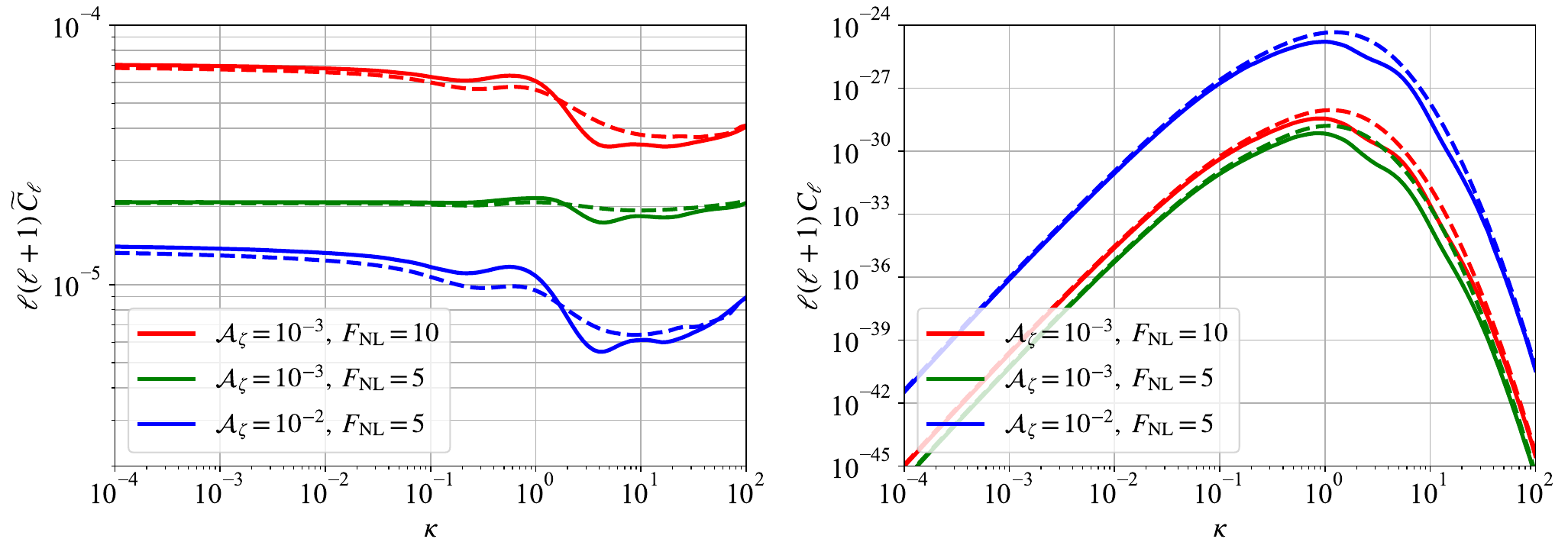}
    \caption{Left panel: Reduced angular power spectra $\ell (\ell+1)\,\widetilde{C}_\ell$ versus the scale $\kappa$ with (solid lines) or without (dashed lines) dissipation. 
    Right panel: Angular power spectra $\ell (\ell+1)\,{C}_\ell$ versus the scale $\kappa$ with (solid lines) or without (dashed lines) dissipation. 
    In both panels, we fix the dissipation parameters to $\alpha=5$, $\gamma=0.3$, and assume a log-normal primordial power spectrum with $\sigma=1$ and $\kappa_\zeta=1$ for $\zeta_g$.}
    \label{fig:Anisotropies}
\end{figure*}

We plot the $\ell (\ell+1)\,\widetilde{C}_\ell(\kappa)$ in the left panel of \cref{fig:Anisotropies}, where the solid (or dashed) lines stand for the results with (or without) dissipation. 
Here, we adopt the same dissipation parameters and $\cP_{\zeta_g}$ as those in \cref{sec:5.1}. 
Dissipation only introduces some small fluctuations as $n_\mathrm{gw}$ in \cref{eq:Cl} is only slightly modified. 
As an important result, we note that the value of $\widetilde{C}_\ell$ is hardly affected by dissipative effects. 
This is because the first term on the right-hand side of \cref{eq:Cl} is largely insensitive to dissipation since dissipation alters the numerator and denominator in a compensating way. 
More physically, this result can be understood by the following fact: dissipative effects only happen inside the horizon, whereas the $\widetilde{C}_\ell$ cares about the physics on large scales. 
As an enlightening perspective, this result suggests that \ac{GW} anisotropies could serve as a robust probe of the primordial Universe (e.g., primordial non-Gaussianity) that is largely immune to subhorizon physics such as dissipation. 
Regarding observations, we further define the angular power spectrum $C_{\ell} = \left[{{\Omega}_{\mathrm{gw},0}}/{(4\pi)}\right]^2\widetilde{C}_{\ell}$. 
Form the right panel of \cref{fig:Anisotropies}, we see that dissipative effects can suppress $C_\ell$ by roughly an order of magnitude around $\kappa\sim\cO(1-10)$. 
This suppression is almost entirely due to the reduction of $\Omega_\mathrm{gw,0}$, making the detection of \ac{GW} anisotropies more challenging at those scales. 

\subsection{Poltergeist mechanism}\label{sec:5.3}
So far, all we consider is the \acp{IGW} in the standard \ac{RD} Universe. 
However, the thermal history of Universe before big bang nucleosynthesis is lack of observational constraints, and many natural cosmological models predict the existence of an \ac{eMD} era (e.g., see Ref.~\cite{Allahverdi:2020bys} for a review). 
If the transition from the \ac{eMD} era to the \ac{RD} era is rapid, \acp{IGW} can be dramatically produced through the ``poltergeist mechanism'' \cite{Inomata:2019ivs,Inomata:2020lmk,Pearce:2023kxp}. 
 predict an \ac{eMD} era, which with a rapid transition to the standard \ac{RD} era.
In this case, \acp{IGW} can be dramatically produced through the ``poltergeist mechanism''.
For simplicity, we assume an instantaneous \ac{eMD}-to-\ac{RD} transition at $\tau_R$. 
In this case, the scale factor and conformal Hubble parameter of the Universe are respectively given by
\begin{align}\label{eq:eMD-RD}
    \frac{a(\tau)}{a(\tau_R)}=
    \left\{
  \begin{aligned}
        & \ ({\tau}/{\tau_R})^2\ ,\ {\tau}\leq\tau_R\ , \\
    & \ {2(\tau-\tau_R/2)}/{\tau_R}\ ,\ {\tau}\geq\tau_R\ ,
  \end{aligned}
    \right.
\mathrm{and}\ 
    \cH(\tau)=
    \left\{
  \begin{aligned}
    & \ 2/\tau\ ,\ {\tau}\leq\tau_R\ , \\
    & \ {1}/{(\tau-\tau_R/2)}\ ,\ {\tau}\geq\tau_R\ .
  \end{aligned}
    \right.
\end{align}
and, the evolution of $\phi_\bk(\tau)$ can be described as 
\begin{align}\label{eq:eMD-RD}
    \phi_\bk(\tau)=
    \left\{
  \begin{aligned}
        & \ \zeta_\bk\times(3/5) \ ,\ {\tau}\leq\tau_R\ , \\
    & \ \zeta_\bk\times(2/3)\,\Phi_\mathrm{RD}(k,\tau)
    \simeq
    \zeta_\bk\times(2/3)\,\Phi^{k_D\rightarrow\infty}_\mathrm{RD}(k,\tau)\,
    e^{-{k^2}/{k_D^2(\tau)}}\ ,\ {\tau}\geq\tau_R\ ,
  \end{aligned}
    \right.
\end{align}
where $\Phi^{k_D\rightarrow\infty}_\mathrm{RD}(k,\tau)$ is determined by the continuity of $\phi$ and $\phi'$ 
\footnote{
The explicit expression of $\Phi^{k_D\rightarrow\infty}_\mathrm{RD}(k,\tau)$ is given by
\begin{equation}
    \Phi^{k_D\rightarrow\infty}_\mathrm{RD}(k,\tau)=
    \frac{9}{10}
    \left\{
    A\,\frac{3\,j_1\left[k(\tau-\tau_R/2)/\sqrt{3}\right]}{k(\tau-\tau_R/2)/\sqrt{3}}
    +
    B\,\frac{3\,y_1\left[k(\tau-\tau_R/2)/\sqrt{3}\right]}{k(\tau-\tau_R/2)/\sqrt{3}}
    \right\}\ ,
\end{equation}
where $j_1(x)$ (or $y_1(x)$) is the first (or second) kind of spherical Bessel function of order 1, and the coefficients $A$ and $B$ are respectively given by \cite{Inomata:2019ivs}
\begin{subequations}
\begin{align}
    &
    A
    =\frac{1}{36} \left\{ 6 \sqrt{3}\, (k\tau_R)\, \sin \left(\frac{k\tau_R}{2 \sqrt{3}}\right)-\left[(k\tau_R)^2-36\right] \cos \left(\frac{k\tau_R}{2 \sqrt{3}}\right)\right\}\ ,
  \\
    &
    B
    =-\frac{1}{36} \left\{\left[(k\tau_R)^2-36\right] \sin \left(\frac{k\tau_R}{2 \sqrt{3}}\right)
    +6 \sqrt{3}\, (k\tau_R)\, \cos \left(\frac{k\tau_R}{2 \sqrt{3}}\right)\right\}\ .
\end{align}
\end{subequations}
}, and we take into account of the dissipative effects in the \ac{RD} era while neglect those in the \ac{eMD} era.
Here, we assume that the damping scale $k_D(\tau)$ follows a power-law scaling for a period of time right after the \ac{eMD} era, parameterized by $\alpha_R$ and $\gamma_R$, i.e.,
\begin{equation}\label{eq:kD_RD}
    {k_D(\tau)}\,\tau_R=
    \gamma_R^{-1}\, \left({\tau}/{\tau_R}\right)^{-\alpha_R/2} \ ,\ \ \tau>\tau_R\ .
\end{equation}
We will focus on the \ac{IGW} production during the \ac{RD} era, which makes a leading contribution. 
Similar to \cref{eq:ogw}, the \ac{GW} spectrum $\Omega_\mathrm{gw}\simeq \Omega_\mathrm{gw,RD}$ can be calculated through the kernel function in the \ac{RD} era, i.e.,
\begin{equation}\label{eq:IIbarRD}
\begin{aligned}
    \overbar{\cI_\mathrm{RD}^2(u,v,k,{\tau})}
    =&
    \int^{k{\tau}}_{k\tau_R} \ud(k{\tau}_1)
    \int^{k{\tau}}_{k\tau_R} \ud(k{\tau}_2)\,
    \times\frac{1}{2}\,
    \bigg\{
    \prod_{i=1,2}
    \big[(k{\tau}_i-k\tau_R/2)\, \cos{(k{\tau}_i)}\,f_\mathrm{RD}(uk,vk,{\tau}_i)\big]
     \\
    &+
    \prod_{i=1,2}
    \big[(k{\tau}_i-k\tau_R/2)\,\sin{(k{\tau}_i)}\,f_\mathrm{RD}(uk,vk,{\tau}_i)\big]
    \bigg\}\  ,
\end{aligned}
\end{equation}
where the source function $f_\mathrm{RD}$ is given by
\begin{equation}\label{eq:fRD}
\begin{aligned}
   f_\mathrm{RD}(uk,vk,{\tau}_i)
   &=\frac{4}{3}\, 
   \Phi_\mathrm{RD}(uk,{\tau}_i)\, \Phi_\mathrm{RD}(vk,{\tau}_i)
    +\frac{4}{9}\, 
   ({\tau}_i-\tau_R/2)^2\,  
   \Phi'_\mathrm{RD}(uk,{\tau}_i)\,  
   \Phi'_\mathrm{RD}(vk,{\tau}_i)
   \\
   &\quad
   +\frac{4}{9}\,
   ({\tau}_i-\tau_R/2)\, 
   \Big[
   \Phi_\mathrm{RD} (uk,{\tau}_i)\, 
   \Phi'_\mathrm{RD}(vk,{\tau}_i)
   +
   \Phi'_\mathrm{RD}(uk,{\tau}_i)\, 
   \Phi_\mathrm{RD} (vk,{\tau}_i)
   \Big]
   \ .
\end{aligned}
\end{equation} 
The ``poltergeist mechanism'' is physically explained as follows: 
For $\phi_\bk$ with $k\tau_R\gg 1$, they keep as constant during the \ac{eMD} era and turn to oscillate very fast (comparing to the cosmic expansion) once the \ac{RD} era begins. 
If the \ac{eMD}-to-\ac{RD} transition is sufficiently rapid, the amplitudes of $\phi_\bk$ have no time to decay during these oscillations. 
Therefore, the \acp{GW} induced by these $\phi_\bk$ right after the \ac{eMD}, estimated as $h\sim \cH^{-2}\phi'\phi' \sim (k\tau_R)^2 \zeta^2 \gg \zeta^2$, would be much more significant than those produced in a standard \ac{RD} Universe, given by $h\sim \zeta^2$. 
Now let us see how the dissipation in the \ac{RD} era affects this mechanism. 

\begin{figure*}[htbp]
    \centering
    \includegraphics[width=1\textwidth]{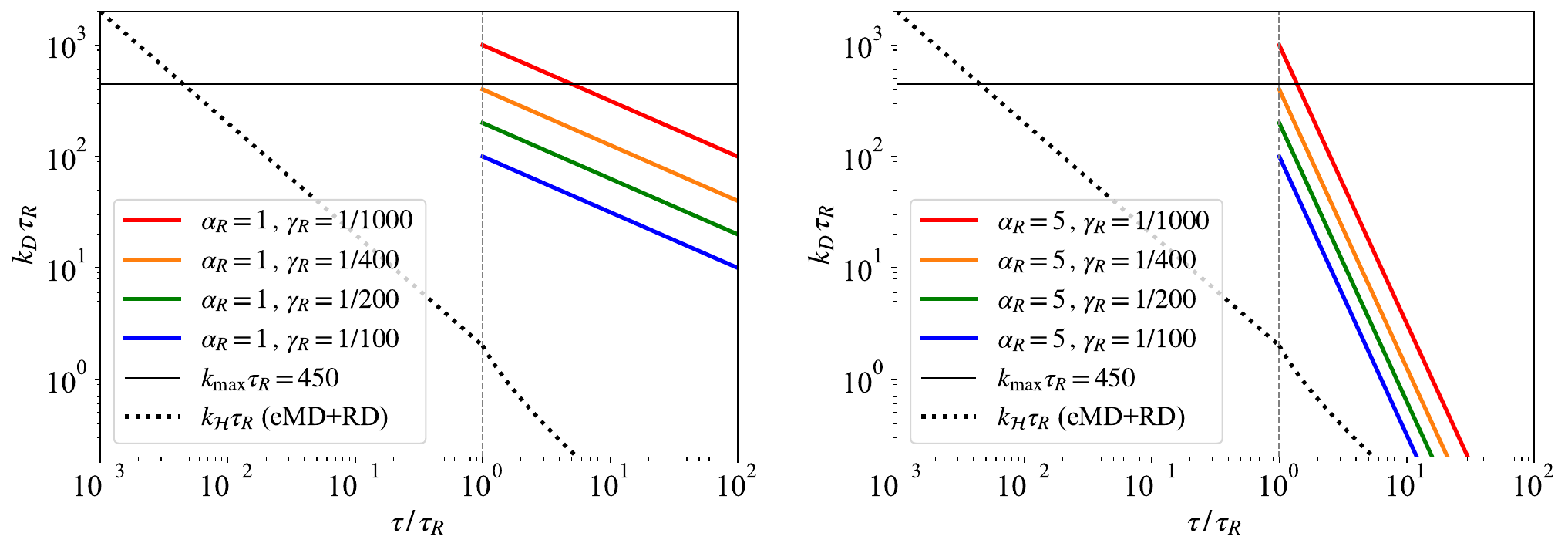}
    \caption{Damping scale $k_D$ right after the \ac{eMD} era (normalized by $1/\tau_R$) as a function of $\tau/\tau_R$ for different values of $\alpha_R$ and $\gamma_R$. The comoving Hubble parameter $\kappa_\cH\tau_R$ and the cutoff scale $k_\mathrm{max}\tau_R$ are also plotted for comparison. 
    In the left (or right) panel, we fix $\alpha_R=1$ (or $\alpha_R=5$) and take $\gamma_R=1/1000, 1/400, 1/200, 1/100$. }
    \label{fig:kD_Poltergeist}
\end{figure*}

\begin{figure*}[htbp]
    \centering
    \includegraphics[width=1\textwidth]{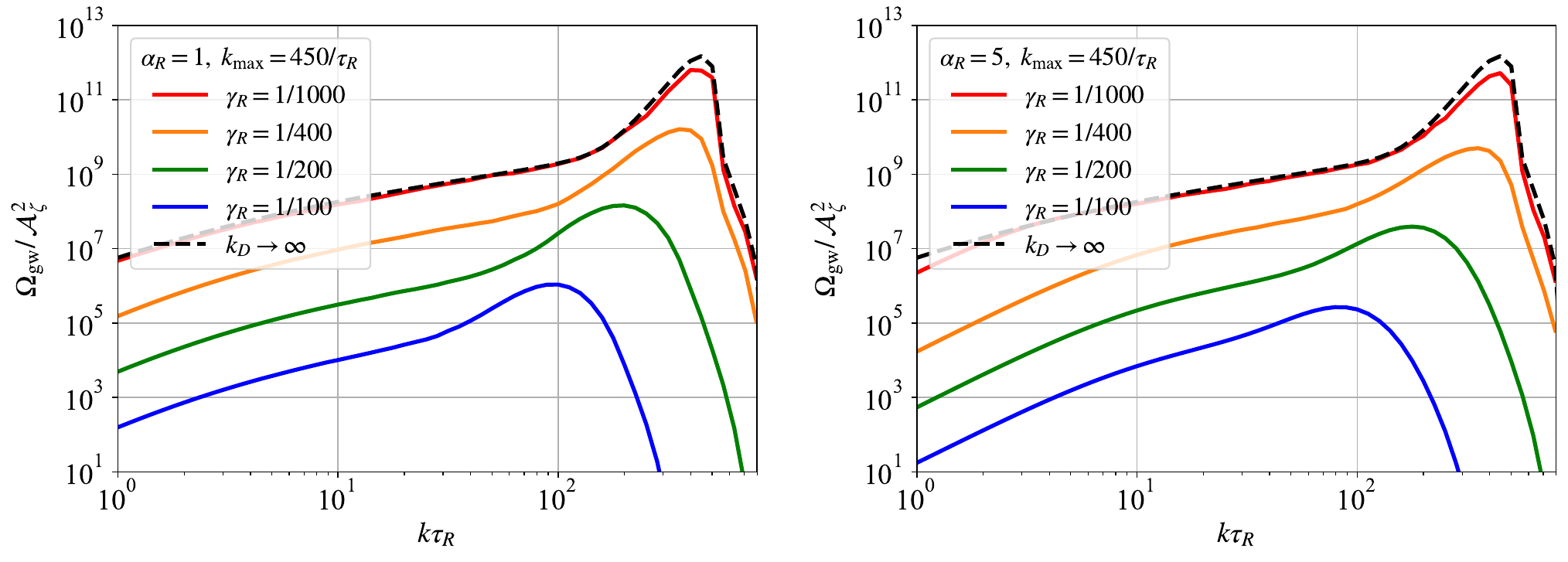}
    \caption{Energy-density spectra of \acp{IGW} (normalized by $\cA_\zeta^2$) in the poltergeist mechanism versus the scale $k\tau_R$ with (solid lines) or without (dashed lines) dissipation.
    In the left (or right) panel, we fix $\alpha_R=1$ (or $\alpha_R=5$) and take $\gamma_R=1/1000, 1/400, 1/200, 1/100$. 
    In both panels, we adopt the primordial curvature power spectrum $\cP_\zeta(k)=\cA_\zeta\,\Theta(k_\mathrm{max}-k)$ with $k_\mathrm{max}=450/\tau_R$. }
    \label{fig:Poltergeist}
\end{figure*}

For different dissipation parameters $\alpha_R$ and $\gamma_R$, we plot the evolution of damping scale $k_D(\tau)$ in \cref{fig:kD_Poltergeist} and the corresponding the \ac{IGW} spectra (normalized by $\cA_\zeta^2$) caused by the poltergeist mechanism in \cref{fig:Poltergeist}. 
Here, we adopt $\cP_\zeta(k)=\cA_\zeta\,\Theta(k_\mathrm{max}-k)$ with the cutoff scale $k_\mathrm{max}= 450/\tau_R$. 
As shown in \cref{fig:Poltergeist}, while the poltergeist mechanism enhances $\Omega_\mathrm{gw}$ to $\sim 10^{12}\cA_\zeta^2$ at $k\sim k_\mathrm{max}$, dissipative effects reduce this enhancement. 
For $\gamma_R<1/(k_\mathrm{max}\tau_R)$, the $\Omega_\mathrm{gw}$ can be reduced by several orders of magnitude compared to the dissipation-free case, and this reduction is highly sensitive to $\gamma_R$.  
The peak position shifts from $k\sim k_\mathrm{max}$ to $k\sim 1/\gamma_R$, which means that on the scales of $1/\gamma_R\lesssim k\lesssim k_\mathrm{max}$, dissipative effects suppress the \ac{IGW} production more than the poltergeist mechanism enhances it. 
However, as long as $\gamma_R>1/(k_\mathrm{max}\tau_R)$, dissipative effects become unimportant and do not change the magnitude of $\Omega_\mathrm{gw}$. 
Furthermore, by comparing the left and right panels of \cref{fig:Poltergeist}, we find that $\alpha_R$ can also slightly change $\Omega_\mathrm{gw}$ (within one order of magnitude). 
These findings can be understood as follows. 
For $k\tau_R\gg1$, poltergeist \acp{GW} are mainly produced in a short period after the beginning of \ac{RD} era ($ k\tau_R \lesssim k\tau \lesssim \cO(1)\times k\tau_R$) \cite{Inomata:2020lmk}, so the dissipative effects almost entirely depend on $k_D(\tau_R)=1/\gamma_R$ and is not sensitive to the subsequent evolution of $k_D(\tau)$. 
Note that the properties of dissipative effects in the poltergeist mechanism are quite different from those in the standard \ac{RD} Universe. 
A notable difference between these two types of \acp{IGW} production is that, the former happens deep inside the horizon (i.e., $k\tau\gtrsim k\tau_R\gg1$), while the latter mainly happens when $\phi_\bk$ just reenters the horizon (i.e., $k\tau\gtrsim1$). 
Therefore, considering that dissipation is ``deep-inside-the-horizon'' physics, its impact on the poltergeist mechanism could be extremely significant, even when $k_D$ is much smaller than $k_\cH$. 
Our results indicate that dissipative effects are an important factor to consider in the poltergeist mechanism.

\section{Conclusions}\label{sec:6}

In the early Universe, due to particle diffusion, the cosmic fluid behaves as a viscous imperfect fluid, inevitably exhibiting dissipative effects.  
These dissipative effects are closely related to the particle interaction in the early Universe and become especially significant near the decoupling of weakly-interacting particles.  
Dissipation can alter the evolution of cosmological scalar perturbations, causing their exponential damping on small scales.  
Since cosmological scalar perturbations induce \acp{GW}, the impact of dissipation on their evolution inside the horizon naturally imprints onto the \ac{IGW} spectrum.  

Dissipative effects can introduce rich phenomenology in \acp{IGW}.  
Dissipation is most prominent around $\kappa \sim 1$, corresponding to the horizon scale at weakly-interacting particle decoupling.  
The detailed phenomenology largely depends on the shape of $\mathcal{P}_\zeta$ near $\kappa \sim 1$, as well as on the dissipation parameters $\alpha$ and $\gamma$.  
For a scale-invariant $\mathcal{P}_\zeta$, dissipative effects produce a characteristic double-valley structure in the \acp{IGW} spectrum near $\kappa \gtrsim 1$, where the positions and depths of the valleys are determined by $\alpha$ and $\gamma$.  
For a monochromatic $\mathcal{P}_\zeta$, dissipative effects smooth the original resonant peak/valley in the \acp{IGW} spectrum and modify the logarithmic infrared behavior, and these changes quantitatively depend on $\alpha$, $\gamma$, and $\kappa_\zeta$.  
For a log-normal $\mathcal{P}_\zeta$, dissipation effects combine the above two cases, featuring significant suppression near $\kappa \sim 1$ and modification of the logarithmic infrared behavior, with properties depending more complexly on $\alpha$, $\gamma$, $\kappa_\zeta$, and $\sigma$.  
All of these phenomenological features can be intuitively explained by introducing the effective ``dissipation-induced'' lifetimes of \ac{IGW} sources.  

Dissipative effects on \acp{IGW} have important implications for future \ac{GW} observations.  
Within the \ac{SM}, dissipation is dominated by neutrinos and substantially affects \acp{IGW} at nanohertz frequencies, an effect that must be considered when analyzing \ac{PTA} data. 
If there exist weakly-interacting particles beyond the \ac{SM}, their dissipation effects may leave distinctive imprints at higher frequencies in the \ac{IGW} spectrum, making them important targets for future \acp{PTA} and space-/ground-based \ac{GW} interferometers.  
In addition, the modification of the logarithmic infrared behavior and the reduction in the total \ac{GW} energy density fraction caused by dissipation are also effects that must be taken into account in future \ac{GW} analysis.

Dissipative effects in \acp{IGW} open an new window for probing new physics.  
The dissipation-induced features in the \ac{GW} spectrum is characteristic imprints of weakly interacting particle decoupling and encode crucial information about their interactions.  
By establishing a general relation between these features (especially their frequencies) and particle interactions in the early Universe, we propose a novel paradigm to search for potential new physics via detecting dissipation effects in \acp{IGW}. 
This provides a new framework to explore new physics at extremely high energy scales, which enhances our capability to explore the early Universe and fundamental physics.  

As an extension, we have also preliminarily discussed the dissipative effects in the presence of primordial non-Gaussianity, as well as its impact on \ac{GW} anisotropies and the poltergeist mechanism.  
When primordial non-Gaussianity is present, the influence of dissipation near $\kappa \sim 1$ would “spread” to other scales due to mode coupling of $\zeta$ at different scales.  
The anisotropies of \acp{IGW} are insensitive to dissipation, as dissipation is essentially a subhorizon effect and does not alter the large-scale distribution of \acp{IGW}.  
The effect of dissipation on the poltergeist mechanism largely depends on the parameter $\gamma_R$.
The poltergeist \acp{GW} can be suppressed by several orders of magnitude if $\gamma_R<1/(k_\mathrm{max}\tau_R)$, reflecting that the importance of considering dissipative effects in this mechanism. 

The study of dissipative effects represents a promising direction for future \ac{GW} cosmology.  
Given the universality of dissipation in the early Universe, it can significantly impact various mechanisms of cosmological gravitational wave production, such as those arising from cosmological phase transitions \cite{Guo:2023koq}.  
Furthermore, since dissipation originates from microscopic particle processes, its associated phenomenology is expected to bear distinctive signatures linked to particle interactions.  
A deeper understanding of dissipative effects not only enables more precise modeling of gravitational wave generation mechanisms but also provides a valuable bridge to deepen our insight into the early Universe and fundamental physics.

\acknowledgments

We would like to express our gratitude to Dr. Jun-Peng Li for helpful discussion. 
This work is supported by the National Natural Science Foundation of China (Grant Nos. 12533001, 12475075, 12175243).








\bibliography{biblio}
\bibliographystyle{JHEP}
\end{document}